\documentstyle[eqsecnum,prb,aps,epsf]{revtex}

\def \T{\mathop{{\rm T}_\tau}\nolimits}
\def \Tr{\mathop{\rm Tr}\nolimits}
\def \Torb{T_K^{\rm orb}}
\def \tens{\underline}

\def \v{{\tens v}}
\def \V{{\tens V}}

\def \t{{\tens t}}

\def \M{{\tens M}}

\def \0{{\tens 0}}
\def \dalpha{{\delta\alpha}}
\def \dv{{\tens {\delta v}}}

\def \etam{{\tens \eta}}
\def \dag{{+}}
\def \exponent{{\Delta}}
\begin{document}
\draft

\title{Orbital Kondo-effect from tunneling impurities}
\author{G.\ Zar\'and}
\address{Institute of Physics, Technical University of Budapest,
P.O.\ Box 112, H-1525, Hungary.}
\author{K.\ Vlad\'ar}
\address{Research Institute for Solid State Physics,
Budapest, P.O.\ Box 49, H-1525, Hungary}
\date{\today}
\maketitle

\begin{abstract}
The article reviews recent results for the low energy
physics of fast tunneling centers in metallic environments.
For strong enough couplings to the environment
these tunneling centers display an orbital Kondo effect
and give rise to a non-Fermi-liquid behavior.
This latter property  is explained by establishing a mapping of the
tunneling center model to the multichannel Kondo model via the
renormalization group transformation combined with a $1/N_f$~expansion.
The case of $M$-state systems, the role of the splittings and 
the present experimental situation are also discussed.
\end{abstract}


\section{Introduction}

Non-Fermi-liquid systems like some heavy fermion materials,\cite{HF}
one-dimensional interacting electrons\cite{1Del} or multichannel quantum
impurity problems \cite{NozBland,CoxZawa} have been the subject of growing
interest during the past few years. 
These systems have the common feature that they display nonanalytical 
power law singularities in various physical properties and that the usual
quasiparticle picture of Landau's Fermi-liquid theory \cite{FLT,KondoFLT}
does not apply for them. 

From the systems mentioned above quantum impurity problems
are of special interest. This is due to the fact that they
provide simple toy models for the study of strongly correlated
lattice models and furthermore the techniques developed for
them can be directly applied to finite \cite{Read,Bickers} and
infinite dimensional lattice
models.\cite{dtoinfty} These latters seem to be very close to the
realistic two or three-dimensional models designed to describe
heavy fermion compounds.\cite{Schiller} Furthermore it has been
suggested that even the properties of high-$T_c$ materials
might be explained by means of multichannel Kondo-like models.\cite{HighT_C}

One of the most interesting quantum impurity problems is provided
by tunneling centers (TC).  These are formed by some heavy
particle (an atom, a point defect or a collective coordinate of
a dislocation) moving in an effective potential
with at least two local minima. If the barrier between these
minima is sufficiently large and the minima are close enough to 
each other then at low temperature the hopping via 
thermal activation is negligible and the atom can move only by
tunneling from one minimum to another.
At low enough temperatures the heavy particle  stays in the
lowest lying quantum states 
associated to the minima of the effective potential and the presence of the
higher excited states is only reflected in the renormalization
of the different coupling constants characterizing the
interaction of the TC with its surrounding (See Sec.~\ref{ss:excited}). 

The physics of an isolated TC is rather boring, but becomes
interesting if the TLS is coupled to its environment. In this paper
we restrict our  considerations to  TC's put into a metal.
A realistic TC in a metal is coupled both to the acoustic phonons and
the conduction electrons providing the low energy excitations of
the environment. However, as also stressed by
Prokof'ev,\cite{Prokofev} since the phonon-TC coupling is
proportional to 
the momentum transfer $q$ and the density of the low-energy
phonon modes scales as $\sim 
\omega^2$ this interaction can be neglected with respect to the
coupling to the low energy electron-hole excitations, which have
a linear density of states and an energy independent coupling
to the TC.

The TC is coupled to the conduction electrons by two different
processes.\cite{KondoTLS,Zawa}
For a fixed position of the heavy particle the conduction electrons tend 
to build up a {\it screening cloud} around the tunneling particle.
Since by a tunneling process the heavy particle has to carry with
itself this  screening cloud consisting of an infinite
number of electron-hole excitations, this interaction tends to
{\it localize} the heavy particle reducing its tunneling rate at low
temperatures. For slow tunneling centers displaying individual
jumps this is the only important process and it has been studied
in great detail both experimentally and theoretically.\cite{Leggett,Kagan_Prokofev,BVZ}
If only this screening interaction is taken into account then the
different vertices occurring in a perturbative expansion commute with 
each other. Therefore this model is sometimes also referred to as the
{\it commutative } model. 

On the other hand, if the tunneling rate is large enough, then
in addition to the screening {\it electron assisted tunneling}
becomes also possible ({\it noncommutative model}).\cite{KondoTLS,VladZaw}
In this process an electron is scattered on the heavy particle while it jumps from one
minimum to another. The latter is essentially due to
barrier fluctuations caused by the local electronic density
fluctuations \cite{VladZaw,ZarSSC} but it can also be generated
by the virtual hoppings to the excited states of the TC.\cite{ZarZaw}
While the amplitude of this process is rather small, it results
in the generation of a Kondo effect and drives the system 
away from the marginally stable commutative model at low energy scales.
As the temperature decreases, the conduction
electrons and the heavy particle form a strongly 
correlated Kondo-like bound state.\cite{Zawa,VladZaw,ZarPRL} For a
two-level system (TLS), the important case of a TC with two
potential minima, the energy of this ground state is
approximately given by the Kondo energy 
\begin{equation}
\Torb = D (v^xv^z)^{1/2} \left({v^x\over 4v^z}\right)^{1/4v^z}\;,
\label{eq:torb}
\end
{equation}
where $D$ is some bandwidth cutoff of the order of the Fermi
energy, and $v^z$ and $v^x$ denote the dimensionless couplings characteristic to screening and assisted
tunneling, respectively. For an $M$-level system no such simple
formula can be given, but the estimated maximal Kondo
temperatures are of the same order of magnitude, $\Torb \sim 1
{\rm K}$.\cite{ZarPRL}

In the considerations of the previous paragraph we have
neglected the role of the {\it splitting} $\Delta$ of the nearly
degenerate levels of the TC. While this splitting is usually
small and it is also renormalized downward during the scaling
procedure it generates a further low energy scale called the
{\it freezing temperature} $T^\ast$. 
Above this scale the levels of the TC can be considered as degenerate.
On the other hand, below $T^*$ the splitting becomes relevant and the internal
degrees of freedom leading to the Kondo effect are in most cases
frozen out.\cite{MF}

Usually one assumes that the interaction of heavy particle and the
conduction electrons is independent of the spin of the
electrons, thus the spin up and spin down conduction electron channels 
are scattered independently and exactly with the same amplitude by the heavy particle.
While the spin is only present as a silent quantum number
(flavor) it plays a crucial role. As discussed later, in the
appropriate temperature range $T^*<T<T_K$ the 
low energy properties of an $M$-state TC can be described by the
two channel $SU(M)\times SU(N_f=2)$ Coqblin-Schrieffer model,
\cite{VladZaw,ZarPRL,ZarVlad,Coqblin} a ***
generalization of the multichannel Kondo model. Here the $SU(M)$
symmetry is connected to the $M$ positions of the TC,
while the $SU(2)$ symmetry is due to the spins of the conduction
electrons. This model, in contrast
to the usual $SU(M)$ single channel model exhibiting a Fermi-liquid
behavior,\cite{NozBland,AndreiFuruya,WiegmanTsvelik} belongs to the family
of overcompensated spin models and displays non-Fermi-liquid
properties due to the $N_f$-fold degeneracy in the 'flavor', and gives
rise to nonanalytical behavior of the different physical
quantities.\cite{ZarPRL,AL_SU(M),Cox_Ruckenstein,Andrei_M} 
In the special case of a TLS ($M=2$) the equivalent model is just the
two-channel Kondo model. While the usual $SU(2)\times SU(2)$
two-channel Kondo model has been investigated by various
techniques such as Bethe Ansatz,
\cite{Andrei,Wiegman,Sacr_Schlott_rev} conformal field 
theory,\cite{AL} large $N$ expansions,\cite{Bickers} numerical
renormalization group,\cite{2CK_NRG} path integral
approach,\cite{VladZimZaw,FabrNozGog} 
and  bosonazitaions,\cite{EK} the number of results
obtained for the $SU(M)\times SU(N)$ model with $M\ge 3$ is
still more restrictive.

In the present review we mainly concentrate on the mapping of the
noncommutative TC model to the $SU(M)\times SU(N)$ model in the
above described temperature range, but we also give a concise
review of the recent experimental and theoretical results in the field.

The paper is organized as follows: In Section~\ref{sec:model}
we shortly discuss the model introduced to describe a tunneling heavy particle.
In Sec.~\ref{sec:ll} we discuss the origin of the logarithmic 
corrections appearing in the perturbation theory and apply
Anderson's 'poor man's scaling' to the TLS
problem.\cite{poorman} Then in Sec.~\ref{sec:nll} we give a
concise  introduction to the multiplicative  renormalization group 
and we show how to obtain the estimation of the Kondo
temperature (\ref{eq:torb}).  Section~\ref{sec:Fixpoint} is devoted
to the development of a large $N_f$ expansion in the 'flavor
degeneracy', $N_f$ for the TLS. We show that up to $1/N_f^2$ order
the generalized $N_f$-flavor TLS model can be mapped to the
$N_f$-channel Kondo model.  The generalization of the above
mapping to the case of $M$-state systems will be discussed in
Sec.~\ref{sec:Mstate}. Some additional results not closely
connected to the subject of this 
paper such as the path integral formalism applied to two-level
systems, the influence  of spin-flip scattering on the TLS, and 
the role of excited states  will be
shortly discussed in Sec.~\ref{sec:new}. We also discuss the
present experimental situation and the possibility of the
observation of the non-Fermi-liquid behavior, predicted in
Sec.~\ref{sec:exp}. Finally, our concluding remarks are given in
Section \ref{sec:conclusions}.

\section{The two-state model}
\label{sec:model}

To investigate the behavior of two-level system (TLS) coupled to
the conduction 
electrons we first have to construct a model that contains all
the relevant ingredients of the realistic situation. In the
present Section we restrict our considerations to the case of 
TLSs but they can easily be extended to the general case of
M-level systems (MLSs).

\begin{figure} 
\begin{center}
\epsfxsize=11cm 
\hskip0.1pt\epsfbox{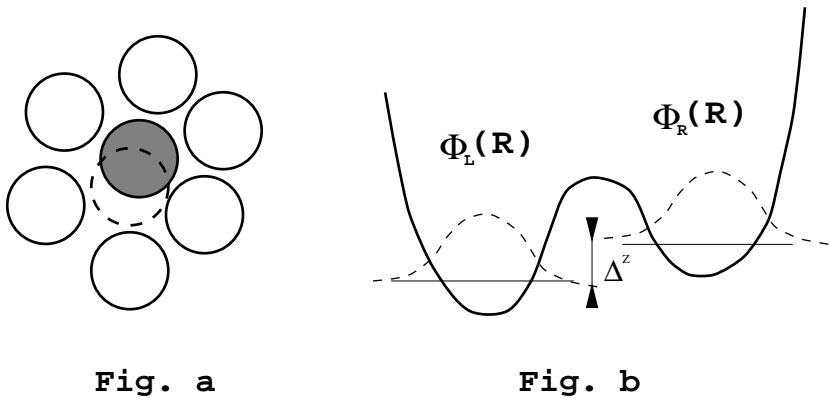}  \\
\end{center}
\vskip0.7truecm  
\caption{\label{fig:TLS} The TLS in an amorphous structure and 
the effective potential describing its motion.} 
\end{figure}

In the nature a variety of physical realizations of TLS's exist.
However, to be specific, constructing the model we consider 
the case of amorphous metals being one of the most frequently
studied systems containing TLSs.\cite{amorphous} The construction
of the model for other systems follows similar lines.
In Fig~\ref{fig:TLS}.a we represent a situation when one or several neighboring 
atoms (ions) in the amorphous structure have two stabile
positions close to each-other.
This situation can be most easily described by some
effective potential shown in Fig.~\ref{fig:TLS}.b, where $R$
denotes the relevant variable in the configuration
space \cite{Leggett} and the two minima are associated to the two
positions of the TLS in Fig.~\ref{fig:TLS}.a. Then one can
associate two 
quantum states $\phi_l(R)$ and $\phi_r(R)$ of the atoms  to the
two minima of the potential well, which
are localized in the  {\it left}- and {\it right}-hand side, respectively.
We assume, that the shape of the double well potential is such
that there are two low-energy states with energies $E_1\approx
E_2$: two linear combinations of $\phi_l(R)$ and $\phi_r(R)$, 
and the next  one has a much higher energy: $E_3-E_1\sim
\omega_{\rm Debye}$. With this assumption the motion of the TLS
at low temperatures will be constrained to the two lowest lying
states and therefore the  states having higher energies can be ignored.
In an amorphous metal the energy distance $\Delta$ between the
two lowest states can differ from TLS to TLS.
In the following considerations we restrict ourselves to a
single TLS and an electron band of the simplest shape. 
To obtain macroscopic quantities we shall treat each TLS as
being independent of the other TLS's and in the end of the
calculation we perform an averaging over the distribution of the
individual TLS parameters. 
The separation of this last summation is reliable at low TLS~concentration, 
i.e., when they are far from one another and do not interact.

At low enough temperatures the Hamiltonian for a TLS could
be described simply by a $2\times 2$ matrix  acting in the
space of the states $\phi_l(R)$ and $\phi_r(R)$.
Nevertheless, for later convenience we rather describe the TLS
motion in terms of an imaginary fermion (pseudofermion),
initially introduced by Abrikosov \cite{Abrikosov} 
to make the second quantized formalism applicable for the Kondo problem.
 In this language the physical states $\phi_l(R)$ and $\phi_r(R)$ 
of the TLS correspond to  the  one particle pseudofermion states
$b^\dag_l|0\rangle$ and $b^\dag_r|0\rangle$, respectively,
$b^\dag_l$ and $b^\dag_r$ being the pseudofermion creation operators.
The nonphysical zero pseudofermion state $|0\rangle$ gives no
contribution to the electron scattering, while the
two pseudofermion state $b^\dag_l b^\dag_r|0\rangle$ can be
projected out by making it thermodynamically very improbable. 
This latter can be achieved if we choose the pseudofermion 
chemical potential $\lambda$ to be very high 
$\lambda\to\infty$. In this limit the leading
terms to any physical quantity will be dominated by the
one pseudofermion states and the two-fermion contributions will 
be suppressed by a factor 
$\sim e^{-\beta\lambda}$ with respect to them. Physical quantities must 
be normalized to the number of pseudofermions present, $
2e^{-\beta\lambda}$. 

The general form of a TLS Hamiltonian can be written as:
\begin{equation}
H_{\rm TLS}=\lambda\sum_\alpha b^\dag_\alpha b_\alpha+
\sum_{i,\alpha,\alpha'}\Delta^i b^\dag_\alpha
\tau^i_{\alpha\alpha'}b_{\alpha'} \;,\label{11}
\end{equation}
the Pauli matrices $\tau^i_{\alpha\alpha'}$ ($i=x,y,z$) denote
the  'pseudospin' of the TLS  and $\alpha=l,r$ labels the
states of the TLS.  
The coefficient $\Delta^z$ describes the asymmetry energy between the
left- and right sides of the TLS while $\Delta^x$ and $\Delta^y$
stand for the tunneling transition (see Fig~\ref{fig:TLS}). 
The splitting of the lowest-lying two states of the TLS can be
expressed via these quantities as $\Delta = 2 (\sum_i
\Delta_i^2)^{1/2}$.

As we shall see, only electrons close to the Fermi surface
interact effectively with the TLS. Therefore,
in the spirit of Landau's Fermi-liquid theory \cite{FLT} the conduction
electrons are treated as noninteracting ones 
and their Hamiltonian is written as
\begin{equation}
H_e=\sum_{\epsilon,n,s}\epsilon a^\dag_{\epsilon ns}a_{\epsilon ns}
\;,\label{12}
\end{equation}
where $a^\dag_{\epsilon ns}$ and $a_{\epsilon ns}$ are the electrons'
creation and annihilation operators with spin $s$ and energy $\epsilon$.
The quantum number $n$ classifies the orbital structure of the states.
It can be, e.g., a spherical wave index-pair $n=(l,m)$ in the
free electron case, 
or a crystal field index reflecting the symmetry of the host.

 In our calculations we mostly use a simplified density of
states for the electrons, 
\begin{equation}
\varrho(\epsilon)=\cases{\varrho_0&if $-D<\epsilon<D$,\cr 0&otherwise,}
\end{equation}
where we assumed that the band can be characterized practically
by only two parameters: the density of states at the Fermi level
$\varrho_0$  and the band width $D$.  The role of the energy dependence of the
density of states will be discussed in Sec.~\ref{sec:new}.

The most general form of the interaction between the electrons
and the TLS  can be written as
\begin{equation}
H_i=\sum_{\textstyle{\epsilon,n,\epsilon^\prime,n',s\atop
\mu,\alpha,\alpha'}} 
a^\dag_{\epsilon ns}b^\dag_\alpha V^\mu_{nn'}\tau^\mu_{\alpha\alpha'}
b_{\alpha'} a_{\epsilon^\prime n's} \;, \label{eq:13} 
\end{equation}
where $\mu={0},x,y,z$ and $\tau^{0}_{\alpha\alpha'}=\delta_{\alpha\alpha'}$.
In the following we use the convention that Greek indices,
$\mu,\nu,..$, take the values $\{0,x,y,z\}$, while the Roman letters,
$i,j,..$,will only be used for the components $x,y,z$.
 Note that in Eq.~(\ref{eq:13}) the couplings $V^\mu_{n
n^\prime}$ are assumed to be energy-independent. This assumption
usually does not influence the infra-read behavior of
logarithmically divergent models. The real electron spin $s$
plays the silent role of {\it flavor}, {\it i.e.},  the
couplings do not depend on it, but its multiplicity will essentially affect
the ground state and the low temperature behavior of the system.

 Assuming a simple two-body interaction between the TLS and
the conduction electrons the coupling constants in
Eq.~(\ref{eq:13}) can be estimated by  simple integrals.\cite{VladZaw,ZarSSC}
The physical interpretation of these coupling constants is as follows.
$V^{0}\pm V^z$ describe the {\it on site} electron scattering on
the two positions of the TLS. 
If the moving particle is an ion different from the host, then
this part of the interaction may be strong, $\varrho_0 V^{0}\pm
\varrho_0 V^z\sim1$. These couplings are responsible for
the generation of the Friedel-oscillations.
The coupling $V^z$, also sometimes called the '{\it screening
interaction}', describes the fact that the electrons 
are scattered differently if the TLS is sitting in its
left- and right position, and if the distance $d$ between the two
sites is small, $k_F d\ll1$, then $V^z\ll V^{0}$. It is the
coupling $V^z$ which is responsible for the eventual
localization of the TLS discussed in the Introduction.

\begin{figure} 
\begin{center}
\epsfxsize=11cm 
\hskip0.1pt
\epsfbox{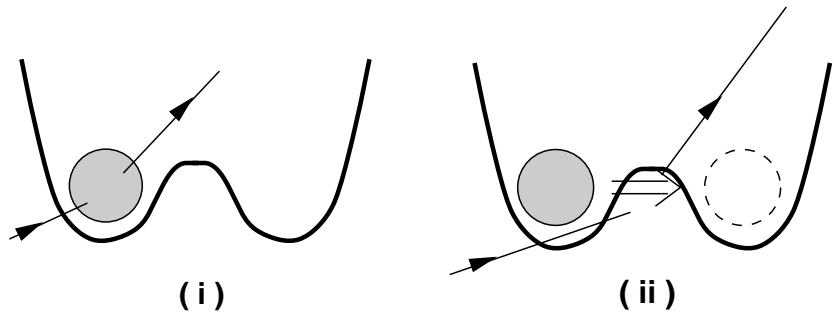}  \\
\end{center}
\vskip0.7truecm  
\caption{\label{fig:interactions} Sketch of the screening
interaction (i) and the assisted tunneling process (ii).} 
\end{figure}

The terms proportional to $V^x\pm iV^y$ describe {\it electron
assisted tunneling} of the heavy particle 
between the two TLS sites, {\it i.e.} processes where a
conduction electron is scattered by the TLS flipping from one
side to the other simultaneously. These processes are
graphically depicted in Fig.~\ref{fig:interactions}. They are
mostly generated by the 
fluctuations of the barrier of the TLS potential due to the
local electronic charge fluctuations.\cite{VladZaw,ZarSSC}
Since they contain the overlap integral between the heavy
particle states $\phi_l(R)$ and $\phi_r(R)$, they are much
smaller then the on site terms: $V^x\pm iV^y\ll
V^z$.\cite{VladZaw} 

\section{Logarithmic corrections in leading order and orbital Kondo effect}
\label{sec:ll}
 Having established our microscopic model we turn to the
analysis of the perturbation series.
We use the Matsubara-technique to handle the interacting 
electron-pseudo\-fermion system.
The following imaginary time Green's functions are introduced for the two
 types of particles:\cite{FLT}
\begin{eqnarray}
&&G_e(\epsilon,n,\tau,\epsilon',n',\tau')=-\bigl\langle\T\bigl[
a_{\epsilon n s}(\tau)a^\dag_{\epsilon'n's}(\tau')\bigr]\bigr\rangle
 \;,\nonumber\\&&
{\cal G}(\alpha,\tau,\alpha',\tau')=-\bigl\langle\T\bigl[
b_\alpha(\tau)b^\dag_{\alpha'}(\tau')\bigr]\bigr\rangle \;.
\end{eqnarray}
Here $\T$ is the $\tau$-ordering operator, the operators
$a_{\epsilon n s}(\tau)$ and $b_\alpha(\tau)$ are in Heisenberg
representation and the angular brackets denote thermodynamic average.
Then the unperturbed Green's functions in Fourier representation
are given by
\begin{eqnarray}
&&G_e^0(\omega)=\bigl(\omega-\epsilon\bigr)^{-1} \;,\nonumber\\
&&{\cal G}^0(\omega)={\textstyle\bigl(\omega- \Delta^i\tau^i
\bigr)}^{-1} \;,\label{eq:30}
\end{eqnarray}
where $\omega$ denotes a Matsubara frequency $\omega=i\omega_n$
and the pseudofermion frequency 
has been measured from the chemical potential $\lambda$.
In Eq.~(\ref{eq:30})
a short matrix notation was used for $G_e$ and ${\cal G}$ 
whose indices $\epsilon n$ and $\alpha$  have been
suppressed, respectively. 

\begin{figure} 
\begin{center}
\epsfxsize=11cm 
\hskip0.1pt
\epsfbox{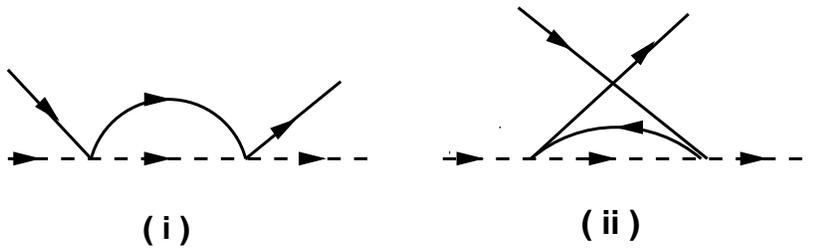}  \\
\end{center}
\vskip0.7truecm  
\caption{\label{fig:2ndvertex} Lowest order vertex corrections
in the perturbation theory.} 
\end{figure}

As in the case of the Kondo effect, the perturbation theories
lead to logarithmically divergent contributions to different
quantities as the relevant energy scale or the temperature tends
to zero.  To demonstrate this we calculate 
the vertex corrections of the lowest order depicted in
Fig.~\ref{fig:2ndvertex},
where the solid lines stand for the electrons and the dotted lines
stand for the pseudofermions. Their contribution is
\begin{eqnarray}
&&\int_{-D}^{+D} \varrho_0 \,d\epsilon\sum_{i,j,n_1,\alpha_1,\alpha_2}\biggl(
V^i_{nn_1}V^j_{n_1n'} \tau^i_{\alpha\alpha_1}
\left({1-n(\epsilon)\over\omega-\epsilon-\Delta^i\tau^i} 
\right)_{\alpha_1\alpha_2} \tau^j_{\alpha_2\alpha'}
\nonumber\\&&\hskip7pc
- V^j_{nn_1} V^i_{n_1n'} \tau^i_{\alpha\alpha_1}
\left({n(\epsilon)\over\omega+\epsilon-\Delta^i\tau^i}
\right)_{\alpha_1\alpha_2} \tau^j_{\alpha_2\alpha'}
\biggr)
\;,\label{31}
\end{eqnarray}
where $\omega$ is the initial energy of the incoming electron,
$\varrho_0$ denotes the density of states at the fermi energy, and 
$n(\epsilon)$ is the Fermi distribution function at
temperature~$T$.  (For the sake of simplicity we put the
pseudofermion energy at the chemical potential $\lambda$: 
$\omega_{\rm ps}=0$.)

The energy integral can be approximated by $\ln(D/\max\{|\omega|,T,\Delta\})$,
where $\Delta=\bigl(\sum_i{\Delta^i}^2\bigr)^{1/2}$, and terms of the order
of unity have been neglected with respect to the logarithm.
For the sake of simplicity we will  consider
the case when $|\omega|$ is the largest energy scale. (The case $T >
\Delta,|\omega|$ can be treated similarly while the discussion
of the case $\Delta > T, |\omega|$ will be posponed to
Sec.~\ref{sec:nll}.) Then the vertex in the first order
approximation (\ref{31}) gets the form: 
\begin{eqnarray}
&&\tens\Gamma^i(\omega)=\V^i-2i\varepsilon^{ijk}\varrho_0\V^j\V^k\ln(D/|\omega|)
+ \ldots
 \label{eq:32}\\&&
\tens\Gamma^0(\omega)=\V^0 \;,
\end{eqnarray}
where $\varepsilon^{ijk}$ is the Levi-Civitta symbol,
the underlined quantities are matrices in the space of electron
orbital states, $V^i_{nn^\prime}\to \V^i$,
and summation must be carried out on the repeated indices.
The logarithmic term occurs only if the commutator
$\bigl[{\tens V}^i,{\tens V}^j\bigr]_-$ does not vanish ({\it
noncommutative model}). In the reality, the three matrices ${\tens V}^x$,
${\tens V}^y$ and ${\tens V}^z$ do not commute for a realistic TLS,
and a commutative model can be relevant only if $\V^x$ and $\V^y$ are 
negligibly small.  We note that the spin ${1\over2}$ Kondo
model is also an example of a noncommutative model, where the
real spin state of the electrons takes over the role 
of the 'orbital spin' $n$.

\begin{figure} 
\begin{center}
\epsfxsize=11cm 
\hskip0.1pt
\epsfbox{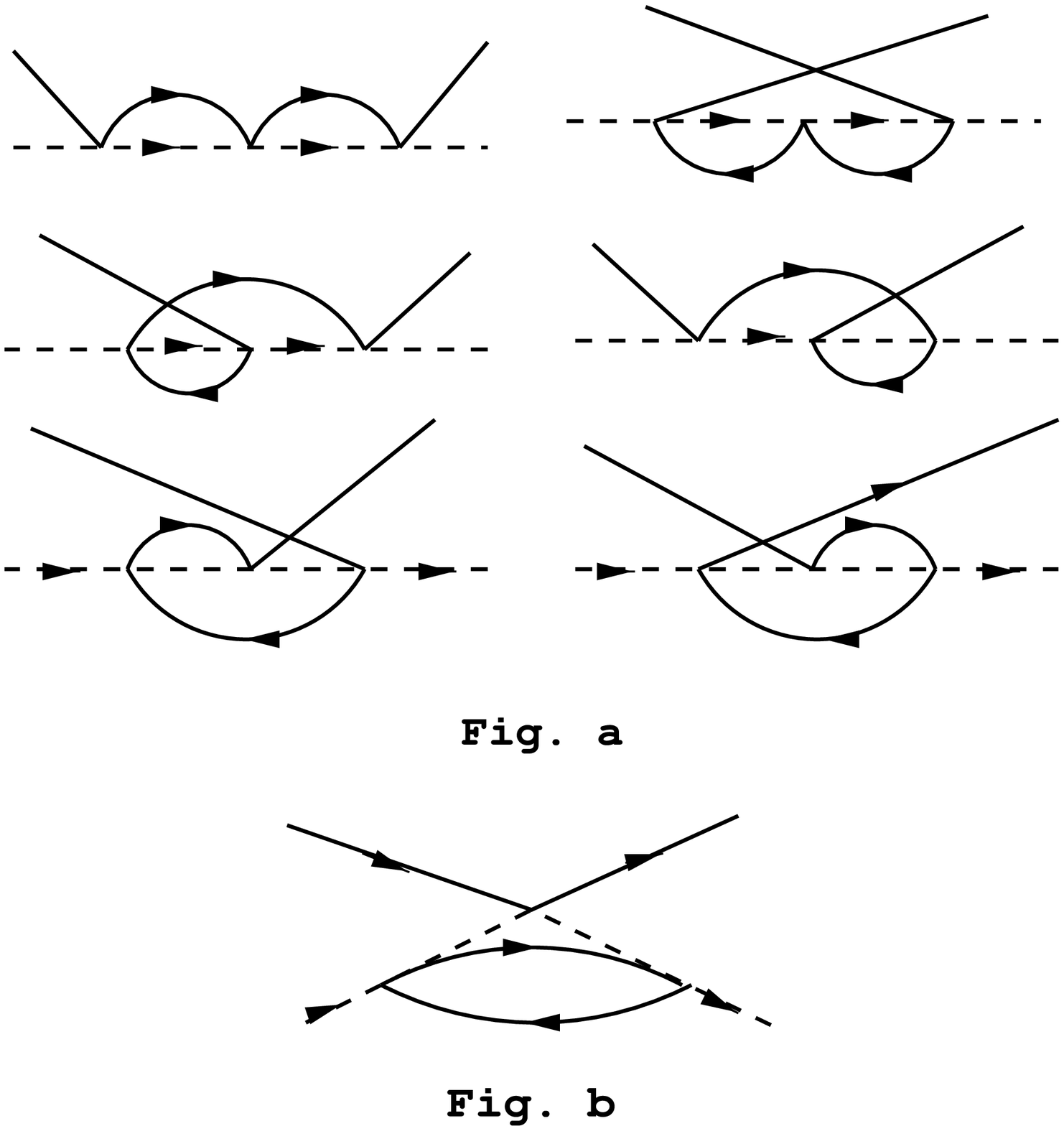}  \\
\end{center}
\vskip0.7truecm  
\caption{\label{fig:third_ord} Second order vertex corrections
generating leading logarithmic (Fig.~a.) and next to leading
logarithmic (Fig.~b.) contributions.} 
\end{figure}

Continuing the calculation of higher order graphs, higher orders 
of logarithms arise: terms proportional to $V^n\ln^{n-m}(D/|\omega|)$,
$m>0$, in general.  According to the usual terminology 
vertex corrections with $m=1$ are called 'leading
logarithmic' while those with $m=2$ are referred to as 'next
to leading logarithmic' diagrams. One can prove that the leading
logarithmic contribution is generated only by the so-called
parquet diagrams -- such vertex diagrams which can be cut in two separate
parts by cutting one electron and one pseudofermion line  in such a way that
the two parts are also parquet diagrams.\cite{Abrikosov}
 In Fig.~\ref{fig:third_ord}.a
we show all the third order leading logarithmic diagrams. 
Up to third order only the diagram in Fig.~\ref{fig:third_ord}.b
gives a next to leading logarithmic contribution. 
Adding the total contribution of the diagrams in
Fig.~\ref{fig:third_ord}.a to the ones in Fig.~\ref{fig:2ndvertex} we
obtain that the vertex function up to third order in the leading
logarithmic approximation can be written as
\begin{eqnarray}
\tens\Gamma^i(\omega)& =
&\V^i-2i\varepsilon^{ijk}\varrho_0\V^j\V^k\ln(D/|\omega|) \nonumber\\ 
&-& 2 \varrho_0^2 \bigl\{ [\V^l,\V^i]\V^l + \V^l[\V^i,\V^l]
\bigr\} \ln^2{D/ |\omega|} + \ldots 
\label{eq:vertexthird}
\end{eqnarray}
At small energies the value of the logarithm can be fairly
large $\ln(D/T)\sim10$ which might
damage the convergence of  the expansion above.
The standard way to make the sum still somehow convergent is  to
sum up an infinite number of diagrams in the perturbation series
that can be achieved by the scaling or renormalization group method.
The '$M$th order scaling equations' will sum all contributions
for which $m\le M$.  In the present and the following  Section
we consider the cases $M=1$ and 2, called as
leading logarithmic and next to the leading logarithmic
approximations, respectively. In Section~\ref{sec:Fixpoint} 
we go even beyond these approximations.

A simple derivation for the leading logarithmic or poor man's scaling 
was given by Anderson for the Kondo problem,\cite{poorman}
which we now apply for the TLS problem.
The scaling hypothesis assumes that there exist  
TLS's having different individual parameters $\V^i$,
$\Delta^i$, and $D$ but having the same low-energy properties,
i.e. they have the same effective TLS-conduction electron
interaction $\Gamma(\omega)$. The parameters of these equivalent
systems are connected by the renormalization group transformation.

The easiest way to construct this transformation is to
consider an infinitesimal transformation of the form
${\V^\prime}^i =\V^i+d\V^i$, $D/D'=1+dx$, and require the
invariance of the vertex function (\ref{eq:vertexthird}) for
arbitrary values of the dynamical variable $\omega$.
As one can check explicitly, the transformation $D\to D^\prime$
and $\V^i \to \V^i - 2 i \epsilon^{ijk} \varrho_0 \V^j \V^k
\ln\left({D\over D^\prime}\right)$ really leaves the expression
(\ref{eq:vertexthird}) invariant. 
The obtained transformation can be cast in the form of the following
differential equation:
\begin{eqnarray}
&&{d\v^i\over dx}=- 2i\varepsilon^{ijk}\v^j\v^k \;,\label{35}\\&&
{d\v^0\over dx}=0 \;,
\end{eqnarray}
where $x=\ln D_0/D$, $D_0$ being the initial bandwidth cutoff,
and the dimensionless couplings $\v^\mu=\varrho_0 \V^\mu$ have
been introduced. Thes 'leading logarithmic scaling equations'
have to be solved with the initial condiction that the couplings
have their bare values at $D=D_0$. The vertex function
(\ref{eq:vertexthird}) can be obtained by integrating these
scaling equations from $x=0$ to $x=\ln{D_0\over \omega}$.

One can check that the integration of Eq.~(\ref{35}) really
reproduces the 
higher order leading logarithmic terms,  and an explicit
summation of the parquet contributions leads to the  same
differential equations,\cite{Abrikosov} i.e., the scaling hypothesis
is correct in leading logarithmic order. We  remark
that the scaling hypothesis is not proved in general for 
its thermodynamic applications, and its validity must
usually be justified term by term in the perturbation series.

As we mentioned in Sec.~\ref{sec:model}, $|\v^{x,y}|\ll |\v^z|$ for a
realistic TLS.    This property makes us possible to
gain some insight to the low-energy properties of a TLS. 
Using the smallness of the assisted tunneling couplings
Eq.~(\ref{35}) can be linearized in $\v^\pm= \v^x \pm i\v^y $
to  obtain
\begin{equation}
{d\v^\pm\over dx}=\pm2\bigl[\v^\pm,\v^z\bigr]_- \;.\label{36}
\end{equation}
and  $\v^z$ is constant in this approximation.
Choosing the electron representation where $\v^z$ is diagonal,
the solution of the previous equations is given by
\begin{equation}
v^\pm_{nn'}(x)=v^\pm_{nn'}(0)\exp\bigl[\pm2x\bigl(
v^z_{n'n'}(0)-v^z_{nn}(0)\bigr)\bigr] \;.
\end{equation}
That matrix element, for which $\pm\bigl(v^z_{n'n'}(0)-v^z_{nn}(0)\bigr)$ 
is the largest, will soon outgrow the others.
This gives a two dimensional subspace where the scaled couplings
get the form of the anizotropic spin ${1\over2}$ Kondo couplings:
\begin{equation}
 v^i_{nn'}=v^i \sigma^i_{nn'} \;,
\end{equation}
$\sigma^i_{nn'}$ being a Pauli matrix acting in the relevant two dimensional 
electronic subspace. From this point the problem continues
similarly to the  anisotropical Kondo problem.\cite{Shiba} 
First the smaller couplings $v^x$ and $v^y$ approach $v^z$, then they
grow further isotropically and diverge at the energy scale
\begin{equation}
T_K^I=D_0\biggl({v^x_0\over4v^z_0}\biggr)^{\textstyle{1\over4v^z_0}} \;,
\label{eq:T_K^I}
\end{equation}
where $v^x_0$ and $v^z_0$ are some characteristic matrix
elements of the unrenormalized couplings.\cite{VladZaw}
$T_K$ is called the Kondo temperature  and its
index $I$ refers to the leading logarithmic approximation.
This divergence is clearly nonphysical, since it would imply
 a finite temperature phase transition in an effectively
one-dimensional system\cite{AL} and, as we shall see later, it is only a
consequence  of the inaccuracy of leading logarithmic approximation.

\section{Next to the leading logarithmic expansion}
\label{sec:nll}

 As we have seen in the previous Section the leading
logarithmic approximation has several discrepancies: It produces
an artificial divergency of the effective couplings at a finite
energy scale, and it does
not tell anything about the scaling of the energy splitting $\Delta$.
As we know from the treatment of the Kondo problem, the leading
logarithmic results can be considerably improved by going one
more logarithmic order further and both the above-mentioned
problems will be solved already at the next to leading
logarithmic level.\cite{Kondo_nll} Therefore in this Section we extend our
previous calculation to the  next to leading logarithmic order.
The summation of the next to leading logarithmic diagrams will
be performed in the framework of the multiplicative
renormalization group.\cite{MRG}

Like in the former Section,
the multiplicative renormalization group scheme provides connection between
equivalent physical systems with different parameters and it is
formulated as an internal symmetry 
of the Green's functions. In our case
the physical systems are characterized by $D$, $\V^\mu$ and also
$\Delta^i$ is involved because it gets perturbative
contributions in this order. 
 The main assumption of the multiplicative renormalization
group is that the Green's functions and vertex functions of the
original and the  scaled TLS's 
have the same functional form, and they only differ by
multiplicative factors $Z_e$ and $Z_p$, which are independent
of~$\omega$:~\cite{Solyom} 
\begin{eqnarray}
&& G_e(\omega/D',V',\Delta')=Z_e(D'/D,V) G_e(\omega/D,V,\Delta)
\;,\label{eq:RG_1}\\&&
{\cal G}(\omega/D',V',\Delta')=Z_p(D'/D,V)
{\cal G}(\omega/D,V,\Delta)\;,\label{eq:RG_2}\\&&
\Gamma^i(\omega/D',V')=Z_e(D'/D,V)^{-1}Z_p(D'/D,V)^{-1}
\Gamma^i(\omega/D,V)\;,\label{eq:RG_3}
\end{eqnarray}
where the primed parameters are those of the renormalized
system, and the  renormalized electron- and
pseudofermion Greens functions are
\begin{eqnarray}
&&G_e=\bigl(\omega-\epsilon-\Sigma_e(\omega)\bigr)^{-1}
\label{eq:G_el} \;,\\
&&{\cal G}={\textstyle\bigl(\omega- \Delta^i\tau^i-
\Sigma(\omega)\bigr)}^{-1} \;.
\label{eq:G_ps}
\end{eqnarray}
In Eqs.~(\ref{eq:RG_1}-\ref{eq:G_ps}) the Green's functions and the
self-energies are matrices just  like in Eq.~(\ref{eq:30}).
 In our case one can easily convince himself that in a given
diagram dressing an internal electron line by a pseudofermion
self-energy results
in subleading corrections in $e^{-\beta\lambda}$. Therefore it
follows that in the $\lambda\to\infty$ limit the electronic wave
function renormalization factor  $Z_e$ is one:\cite{VladZaw}
\begin{equation}
Z_e = 1 \;.
\end{equation}

As in the previous Section the infinitesimal renormalization group transformations
(\ref{eq:RG_1}-\ref{eq:RG_3}) can be written as differential equations
for the different couplings of the model.
The most general form of the scaling equation for the couplings $v$ is
\begin{equation}
{dv\over dx}=\beta(v)\quad,\label{52}
\end{equation}
where $\beta(v)$ is a polynomial, which is determined by
perturbation theory, and the different couplings are represented by a 
single symbol~$v$. 
In the leading logarithmic approximation the polynomial 
is of second order, and it can be determined by taking the
derivative of the vertex (\ref{31}) with respect to $\ln D$.
In this case Eq.~(\ref{52}) generates the leading logarithmic
terms to all orders. To get the next, third order term in
$\beta(v)$ the vertex and pseudofermion self-energy corrections
must be treated simultaneously.  

\begin{figure} 
\begin{center}
\epsfxsize=11cm 
\hskip0.1pt
\epsfbox{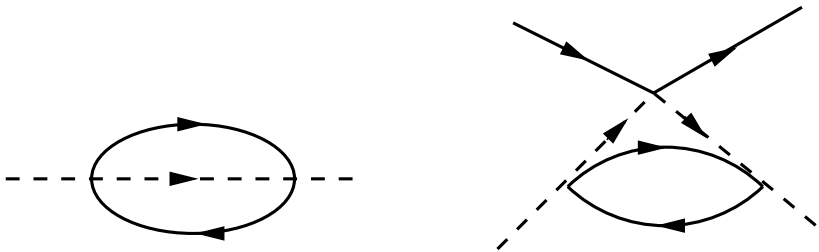}  \\
\end{center}
\vskip0.7truecm  
\caption{\label{fig:nll_gen} Diagrams generating the next to
leading logarithmic contribution to the scaling equations.} 
\end{figure} 

The get some insight what the physical meaning of the scaled
couplings is, let us assume for a moment that the temperature is
the largest energy variable of the vertex function (i.e., we
investigate scattering of electrons {\it at} the Fermi surface
off the impurity), and apply the renormalization group
transformation with the special choice $D^\prime =T$. Then,
since $\ln {D^\prime \over T}\approx 0$ no vertex corrections
appear for the scaled TLS with the primed parameters. Therefore
${\tens\Gamma}^i (T,D',V')\approx {\V^\prime}^i$. Thus 
the solution of Eq.~(\ref{52}) at the scale $x=\ln{D\over T}$,
up to the numerical factor $Z_p$, is just the 
effective interaction of the TLS and the conduction electrons
at temperature $T$. (Knowing the solution of Eq.~(\ref{52}) one
can easily determine the factor $Z_p$ as well). Similar
procedures can be applied to calculate physical quantities like
the electronic scattering rate or the impurity specific
heat.\cite{ZarVlad,Gan}

Calculating the vertex and pseudofermion self-energy
diagrams shown in Fig.~\ref{fig:nll_gen}, we obtain
\begin{eqnarray}
&&\tens\gamma^i=\v^i+\bigl\{-2i\varepsilon^{ijk}(\v^j\v^k)
+N_f[2\v^j\Tr(\v^i\v^j)-\v^i\Tr(\v^j\v^j)]\bigr\}\ln(D/|\omega|)
\;,\label{eq:vertex}\\&&
\Sigma(\omega)=-N_f\Tr(\v^i\v^j)\tau^i\bigl[ \omega I
- \Delta^k\tau^k\bigr]\tau^j\ln(D/|\omega|)
\;,\label{eq:Sigma}
\end{eqnarray}
where $\gamma=\varrho_0\Gamma$ is the dimensionless vertex,
the trace operator ${\rm Tr}\{...\}$ is acting in the 
electronic indices, and $N_f$ is the spin degeneracy of the
conduction electrons ($N_f=2$ for the physical case).  We remind the
reader that underlined quantities denote matrices in the 
orbital indices of the electrons.
The self-energy contains terms proportional to $\omega\ln{D\over
\omega}$, which give a contribution to~$Z_p$.

To construct the scaling equations, an infinitesimal step $D/D'=1+dx$ 
$\bigl(x=\ln(D_0/D)\bigr)$ must be considered in
Eqs.~(\ref{eq:RG_1}--\ref{eq:RG_3}), (\ref{eq:G_ps}),
(\ref{eq:vertex}) and (\ref{eq:Sigma}).  
 Then plugging Eq.~(\ref{eq:Sigma})
into Eq.~(\ref{eq:G_ps})  the renormalization factor $Z_p$ 
can easily be determined by comparing the $\sim1/\omega$ terms
on the two sides of Eq.~(\ref{eq:RG_2}) :
\begin{equation}
Z_p(D'/D,V')=1-N_f\ln(D'/D)\Tr(\v^i\v^i) \;.\label{eq:Z_p}
\end{equation}
Knowing $Z_p$ one can also read off the
infinitesimal transformations of $\Delta^i$ from 
Eq.~(\ref{eq:RG_2})  and (\ref{eq:Sigma}):
\begin{equation}
{\Delta'}^i = \Delta^i+2N_f\ln(D'/D)\bigl[\Delta^i\Tr(\v^j\v^j)
-\Delta^j\Tr(\v^i\v^j)\bigr]+{\cal O}(v^3) \;,\label{5.11}
\end{equation}
which can be written in a differential form as
\begin{equation}
{d\,\Delta^i\over dx}=-2N_f\bigl[\Delta^i\Tr(\v^j\v^j)-\Delta^j
\Tr(\v^i\v^j)\bigr]  \;.\label{5.14}
\end{equation}
This equation describes the renormalization of the TLS splitting
as a function of the temperature. One can easily prove that 
the $\Delta^i$'s are always decreased under the scaling transformation.

The scaling equations for the couplings $\v^i$ can be
generated in the same way using Eqs.~(\ref{eq:Z_p}),
(\ref{eq:vertex}), and (\ref{eq:RG_3}):
\begin{equation}
{d\v^i\over dx}=-2i\varepsilon^{ijk}\v^j\v^k
-2N_f\bigl[\v^i\Tr(\v^j\v^j)-\v^j\Tr(\v^j\v^i)\bigr] \;.\label{5.13}
\end{equation}
These scaling equations can be solved with some realistic restrictions on
the initial couplings~$\v^i$ and it can be shown just like in the
leading logarithmic case that the relevant electron subspace is
two-dimensional  in the weak coupling regime of the scaling.\cite{VladZaw}
As we will show in Sec.~\ref{sec:Fixpoint} this statement remains
valid even below $T_K$ and the only stable fixed point of
Eq.~(\ref{5.13}) is where the interaction has the same structure
as the spin ${1\over 2}$ Kondo coupling, $v_{nn'}^i \sim 
\sigma^i_{nn'}$. 

\begin{figure} 
\begin{center}
\epsfxsize=7cm 
\hskip0.1pt
\epsfbox{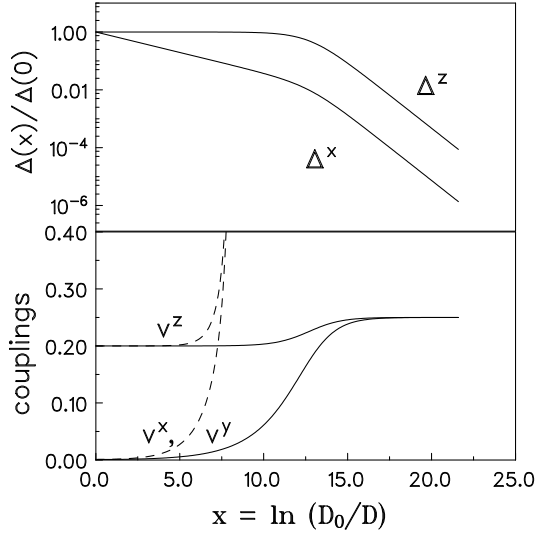}  \\
\end{center}
\vskip0.7truecm  
\caption{\label{fig:Vladscaling} Scaling of the dimensionless
couplings and the splittings for a simplified TLS with $\v^i =
v^i \tens{\sigma}^i$ in the leading- (dashed lines) and next to leading 
logarithmic (continuous lines) approximation.} 
\end{figure}

Typical scaling curves are given in Fig.~\ref{fig:Vladscaling},
for the special 
case where the $\v^i$'s are proportional to some Pauli matrix
$\v^i = v^i \tens{\sigma}^i$ (no summation). As it is shown in
Ref.~\onlinecite{VladZaw} this is a fairly good approximation for a
 TLS interacting with a free electron band.
Both  the leading logarithmic and the next to leading
logarithmic scaling trajectories are presented.
As one can see from the Figure, due to the presence of the
third order terms in the scaling 
equations the couplings $\v^i$ do not diverge anymore, but they
scale to a finite value. 

Then the orbital Kondo temperature is identified by the
characteristic energy scale, where the couplings approach their
fixed point value, and it can be expressed as 
\begin{equation}
T_K^{II}=D_0\bigl(v_0^x v_0^z \bigr)^{N_f/4} \biggl({v_0^x \over4v_0^z }
\biggr)^{\textstyle{1\over4v_0^z }}
 \;,\label{5.16}
\end{equation}
where the indices $II$ and $0$ refer to the next to leading logarithmic
approximation and the initial value of the parameters,
respectively. This Kondo scale like is {\it invariant under scaling}.

Due to the appearance of the prefactor this Kondo
temperature is considerably smaller than the one obtained in the
leading logarithmic approximation, Eq.~(\ref{eq:T_K^I}). While for
reasonable parameters $T_K^I$ is of the order of $\sim 10 K$
in the next to leading logarithmic approximation $T_K^{II}\sim
1K-0.1 K$.

To close this Section we estimate the renormalization of
the tunneling amplitude $\Delta^x$. To be specific we consider
a symmetrical TLS with  $\Delta^z=\Delta^y=0$, and the
above-mentioned simplified couplings $\v^i = v^i \tens{\sigma}^i$.
Then the scaling equation for $\Delta^x$ can be written as
\begin{equation}
{d\,\ln\Delta^x\over dx}= - 4N_f\bigl[(v^y)^2+(v^z)^2\bigr]
\;.\label{eq:Delta^x} 
\end{equation}
The scaling procedure should be stopped by~$\Delta^x$ for very
small~$x$,  where the splitting becomes 
the dominating low-energy scale in the argument of the
logarithms (see the discussion below Eq.~(\ref{31})).
The  corresponding energy scale is defined by the implicit
equation $\Delta^x(\ln(D_0/T^*))=T^*$. 
 Below this energy scale the motion of the TLS is frozen out
 and $\Delta^x$ remains constant for $D < T^*$.  Taking into
account that in most of the scaling procedure $v^z\approx
v^z_0\gg v^x$ in Eq.~(\ref{eq:Delta^x}), the 'freezing
temperature' $T^*$ can be roughly estimated as 
\begin{equation}
T^*=\Delta_0\biggl({\Delta_0\over D_0}\biggr)^
{\textstyle{2N_f{v^z_0}^2\over1-2N_f{v^z_0}^2}} \;.\label{5.20}
\end{equation}
 $T^*$ is sometimes also referred to as the renormalized or
effective splitting of the TLS.\cite{Kagan_Prokofev} For realistic parameters
$T^*/\Delta^x_0$ can be as small as $10^{-2}$.
Taking into account, however, that in the noncommutative model
$v^z$ and $v^x$ also depend on $x$ one obtains that $\Delta^x$ can be
reduced by several orders of magnitude
for large enough $v^z$'s as can be seen in
Fig.~\ref{fig:Vladscaling}.  A more detailed discussion shows
that the renormalization of $\Delta^z$ is 
usually much smaller, and the more symmetric the potential 
the larger is the reduction of energy splitting.

\section{Stability of the two-channel Kondo fixed point: $1/N_F$
expansion}
\label{sec:Fixpoint}

As we have seen in the previous  Sections the
noncommutative TLS shows a strongly correlated behavior at low
temperature. This behavior manifests in the appearance of
logarithmic singularities in the vertex function which can be
associated to the formation of a Kondo-like ground state with a
binding energy of the order of $T_K$.  Using the multiplicative
renormalization group method we were 
able to handle these logarithmic singularities.
In the next to leading logarithmic
approximation the artificial divergence of the vertex function
has been removed and we could also account for the
renormalization of the TLS splittings. 
However, for small values of $N_f$ (including the physical case
$N_f=2$) the next to leading logarithmic approximation breaks
down in the vicinity of the fixed point, 
as the higher order terms of the $\beta$ function in Eq.~(\ref{52}) 
become comparable with the first  ones. The main purpose
of  the present Section is to go below the Kondo scale, and see
what happens to the TLS  at energy scales  $T,\omega<T_K$.
We shall circumvent the above-mentioned  difficulty  by making a $1/N_f$
expansion. 




We have seen in the previous Sections that at high
temperatures there are two orbital electron channels which
dominate the scattering. The logarithmic anomalies are essentially due to
these two channels and the other orbital channels give only a
small contribution. We have also shown that if only these
two channels are considered, then in the weak coupling region
the (now $2\times 2$) matrices $\v^i$ scale towards
simple spin 1/2 operators $\v^i \sim {\tens\sigma}^i=2S_e^i$ with 
$S_e=1/2$.\cite{Zawa,VladZaw} Gided by these
facts it has been conjectured in Ref.~\onlinecite{VladZaw} that at the low
energy fixed point the electron-TLS interaction is described by
a simple effective exchange interaction in the {\it orbital}
degrees of freedom $\sim \tau^i S_e^i$, and thus the model is
equivalent to that of the 
two-channel spin Kondo problem \cite{NozBland} which 
exhibits non-Fermi liquid properties.  The
double degeneracy of the two 'channels' in the latter model
corresponds to the {\it real spin} indices of the conduction electrons
in the TLS case, and the splitting $\Delta$ of the TLS acts as a
local magnetic field at the imputity site in the two-channel
Kondo analogy (See also Table~I an Sec.~\ref{sec:Fixpoint}).
Several recent experiments on metallic point
contacts\cite{Ralph_Buhr,Ralph_Ludwig,Jan_exp} have been
interpreted in terms of the {\it complete equivalence} of the
aforementioned two models (See also Sec.~\ref{sec:exp}). 
This 'complete equivalence' is, however, far from being trivial.
As indicated above, the mentioned results of Vlad\'ar and
Zawadowski are only trustworthy in the high-energy (weak
coupling) region where the renormalized couplings are small, and
their approximations loose their meaning in the strong-coupling
region $T, \omega\sim T_K$.

Vlad\'ar and Zawadowski have already remarked that the 
TLS model has other fixed points which differ from the previously 
mentioned one by the {\it number of orbital channels coupled} to
the TLS. An example of such fixed points is given by $\v^i\sim
S^i_e$ with $S_e > 1/2$. 
Simple estimations show that the coupling of the different
orbital electron channels is strong thus one would
naively expect that as soon as the couplings of the two orbital channels
dominant at $T\gg T_K$ become of the order of unity all the
other couplings start growing up as well. Therefore, considering
for instance three different orbital channels one could imagine  that
although there are two dominant channels at high temperatures
($T>T_K$), the third channel becomes also important below the
Kondo temperature $T_K$, and the low temperature scattering is 
described by an $S_e=1$ orbital electron spin corresponding to
the $2S_e+1=3$ orbital channels. Thus, naively,
one could expect a series of Kondo effects corresponding to the
increase of the couplings of the different orbital channels. 
Vlad\'ar and Zawadowski have also remarked already in their
early work\cite{VladZaw} that below $T_K$ higher angular
momentum scattering might be relevant.

Furthermore, as it is wellknown from renormalization group
theory,\cite{Wilson} the low temperature behavior of a model is
not only determined by the structure of the stable fixed points
it scales to, but also by their {\it operator content}.
Therefore, to identify two models one must
be very careful and has to consider the operator content of the
two models as well. There
are several models, which possess a simple fixed point, but have
nontrivial operator content.  An example is the fixed point
found by H.\ Pang for a generalized two channel Kondo
problem \cite{2CK_NRG} which has a spectrum composed from two
independent Fermi liquid spectra, however, its thermodynamical
properties are claimed to be determined by a non-Fermi-liquid leading
irrelevant operator.  

Our purpose in this Section is to establish a more rigorous
correspondence between the two models and to show that the
stable low-energy fixed point of the model is correctly
described by the conduction electron orbital spin $S_e=1/2$. 
For this purpose we investigate an $N_f$-flavor TLS
Hamiltonian (\ref{12}) and (\ref{eq:13}) with a general spin
degeneracy $N_f$ and develop a systematic $1/N_f$ expansion.
Then we classify all the possible fixed 
points of the model and investigate their stability and operator
content. We find, that the only stable fixed point is 
the $S_e=1/2$ fixed point independently of $N_f$ and the
number of orbital channels considered. While we find that the
operator content of the $N_f$-flavor TLS model is much richer than that of
the $N_f$-channel spin 1/2 Kondo model, the dimension of the
leading irrelevant operators, and thus the thermodynamical and
dynamical behavior of the two models is qualitatively the same. 
Since our result are exact in the limit where $N_f$ is large
and they are independent of the special value of $N_f$ we
expect that these results are also valid even for the  $N_f=2$ case.

In these considerations we first neglect the role of the
splitting of the TLS which serves as a lower cotoff in the
scaling procedure. The role of this splitting will be discussed
later.  First we describe the calculation in the $1/N_f$ order.
Then we discuss the technicalities of the extension of the
calculation to $1/N_f^2$ order, and finally we exploit the
mapping found to describe the non-Fermi liquid properties of the
TLS in terms of the 2-channel Kondo model.

\subsection{The $1/N_f$ order analysis of the scaling equations}

The next to leading logarithmic scaling equations 
for the $N_f$-flavor TLS model have been derived in Sec.~\ref{sec:nll}:
\begin{equation}
{d\v^i \over dx} = -2i \epsilon^{ijk}\v^j \v^k - 2N_f
\v^i \Tr \left\{ \v^j \v^j \right\}
+ 2N_f \v^j \Tr \left\{ \v^i \v^j \right\}\;.
\label{eq:N_f:nllscaling}
\end{equation}
We remind the reader that the $N_f$ factor appearing in the last
two terms of Eq.~(\ref{eq:N_f:nllscaling}) are due to the
presence of the electron loops in the third order vertex
correction and the pseudofermion self energy in Fig.~\ref{fig:nll_gen}. 
We stress that the scaling equation
(\ref{eq:N_f:nllscaling}) is very general. To derive it one has
to assume only that some sort of Fermi surface exists and that
the conduction electrons can be described in terms of the Fermi
liquid theory. The special choice of the density of states has
no effect on the universal properties of the system up to this
order.

The fixed points of Eq.(\ref{eq:N_f:nllscaling}) are determined by
the condition that its right-hand side vanishes. From the
presence of the factor $N_f$ immediately follows that the fixed
point couplins of the TLS should be of the order of $\sim
1/N_f$.  Therefore, as observed first by Nozi\`eres and Blandin for the
overscreened Kondo model,\cite{NozBland} in the $N_f \to \infty$ limit the
scaled couplings remain in the small coupling region and the
scaling equations give exact results. (This should be
contrasted to the $N_f=1$ 'simple Kondo' case where it can be
shown with other methods that the
couplings scale to infinity.\cite{KondoFLT}) 

The fact that the fixed point coupling is of the order of
$1/N_f$ makes also possible to develop a systematic $1/N_f$ 
expansion for our $N_f$-flavor TLS
model.\cite{Muram_Guinea,Gan,ZarVlad} To 
understand this point let us investigate an $n$'th order logarithmic
vertex correction in the perturbation series. Around the fixed
point each vertex is of 
the order $1/N_f$ which results in a factor $1/N_f^n$.  This
small factor is partially compensated by the 
electron loops present in the diagram bringing up a factor
$N_f^L$, where $L$ denotes the number of electron loops in the
diagram. Since this last number is always smaller than $n/2$ 
up to a given order in $1/N_f$ there exist only a
finite number of diagrams which may generate corrections to the
scaling equations around the fixed point. This analysis can be
extended to terms generated by the pseudofermion self-energy
corrections. Therefore, if one wants to calculate the fixed point
coupling or the scaling exponent of the different operators up
to a given order in $1/N_f$ only a finite number of diagrams
must be considered (see also Subsection~\ref{ss:1/n_f^2}). A quick
analysis shows that the lowest order diagrams in $1/N_f$ are
exactly the ones generating the next to leading logarithmic
scaling equations shown in Fig.~\ref{fig:nll_gen}.

Now we turn to the analysis of the classification of the fixed
points of Eq.~(\ref{eq:N_f:nllscaling}). It is easy to
show that the last term in Eq.(\ref{eq:N_f:nllscaling}) can be
eliminated from the fixed point equation by making an orthogonal
transformation $\v^i \rightarrow \sum_j O_{ij}\v^j$, $O$ being
an orthogonal matrix. Therefore, it is enough to consider the
first two terms on the right-hand side of
Eq.(\ref{eq:N_f:nllscaling}):
\begin{equation}
\sum_{j,k} \epsilon^{ijk}\v^j \v^k = i N_f \v^i \sum_{j\not= i}
Tr\left\{ \v^j\v^j \right\} 
\label{eq:N_f:fp_eq}
\; .
\end{equation}
Multiplying Eq.(\ref{eq:N_f:fp_eq}) by $\v^i$ and taking its
trace one  obtains the following equations:
\begin{equation}
iN_f \alpha_i \bigl(\sum_{j\not= i} \alpha_j\bigr)  = \beta 
\; ,\phantom{nnn}(i=x,y,z)\; ,
\label{eq:N_f:alpha_i}
\end{equation}
where $\alpha_i = Tr\{\v^i \v^i \}$  ($i=x,y,z$) and 
$\beta = Tr\{\v^x\v^y \v^z - \v^z
\v^y\v^x\} $. From Eq.(\ref{eq:N_f:alpha_i}) immediately follows
that either at least two of the $\alpha_i$'s are zero at the
fixed point or they are all equal: $\alpha_x = \alpha_y = \alpha_z = \alpha$.
The first case corresponds to the commutative TLS, where
the assisted tunneling is ignored and the couplings $\v^x$ and
$\v^y$ identically vanish.  This is evidently 
an unstable fixed point.  In the second case it is worth introducing
the matrices $J^i = {1\over 2 N_f \alpha} \v^i$.
Then Eq.(\ref{eq:N_f:fp_eq}) tells us that the $J^i$'s
satisfy the SU(2) Lie algebra 
\begin{equation}
[J^i,J^j] = i \epsilon^{ijk} J^k \; .
\label{eq:N_f:SU(2)}
\end{equation}
Therefore the general form of the $J^i$'s at 
the fixed point can be given by a direct sum of finite
dimensional irreducible representations of the SU(2) spin
algebra 
\begin{equation}
J^i = \bigoplus\limits_{k=1}^n S^i_{(k)}\; ,
\label{eq:N_f:directsum}
\end{equation}
where the $S_{(k)}$'s denote integer or half-integer spin
representations and $n$ is the number of irreducible
representations involved. The $S^i_{(k)}$'s in
Eq.(\ref{eq:N_f:directsum}) are acting only in 
a finite dimensional subspace of the total electronic phase
space, and in the rest of the phase space the $J^i$'s give
identically zero. The value of $\alpha$ can easily be determined
using its  definition: 
\begin{equation}
\alpha ={3 \over 4 N_f^2 \sum_k S_{(k)}(S_{(k)} + 1)(2 S_{(k)} +
1)}\; .
\label{eq:N_f:alpha}
\end{equation}
It is obvious from Eq.(\ref{eq:N_f:alpha}) that the value of
$\alpha$ depends in an essential way on the electronic orbital
spin structure of the fixed point approached by the $\v^i$'s. 
Carrying out the scaling all the $\alpha_i$'s become equal and
they scale to a special value of $\alpha$ characteristic to the 
fixed point.

The real low-energy properties of a TLS can only be described by
some {\it stable} fixed points. The presence of the
other unstable fixed points could only be observed if the
paramaters of the TLS are finetuned, but they cannot produce a
universal scaling of the different physical quantities.\cite{Andrei_crossover}
Therefore, having found all the fixed points of Eq.~(\ref{eq:N_f:nllscaling}),
we turn to the stability analysis of the fixed points. 

\vskip0.4truecm
{ \noindent \it 1. The unstable  fixed points}
\vskip0.4truecm

We first show that the
only possible stable fixed point is the one which is equivalent
to a single $S_e=1/2$ orbital electron spin. The proof proceeds in two steps.
First we show that the 'composite' fixed points where the
representation $J^i$ is reducible and contains several
different spin representations are unstable. In the second step
we prove that the fixed points where the $J^i$'s are
irreducible but are equivalent to an $S_e>1/2$ spin representation
are also unstable. Therefore, one can conclude that only that
fixed point can be stable where the couplings $\v^i$ 
contain a single $S_e=1/2$ spin representation.  The stability
of this latter will be proved in the next Subsection.

To prove the instability of a fixed point our strategy  is  to
find a single unstable scaling trajectory running out from the
fixed point. If the $\v^i$'s are composed from at least two spin
representations we 
can chose two of them, $S_{(1)}$ and $S_{(2)}$, and consider
small deviations from the fixed point $\v^i = (2N_f\alpha)
\sum_k S_{(k)}^i$ of the form 
\begin{equation}
\delta\v^i = 2N_f(\delta\alpha_1 S^i_{(1)}
+\delta\alpha_2 S^i_{(2)}) \; . 
\nonumber
\end{equation}
These special scaling trajectories lie in the subspace
expanded by the two arbitrarily chosen spin representations
$S_{(1)}$ and $S_{(2)}$. Linearizing the scaling equations one
obtains a closed system of equations for the $\delta
\alpha_i$'s which can be solved easily. Then it is trivial to show
that the operator 
\begin{equation}
\delta\v^i \sim \left(\alpha_1 S^i_{(1)} -\alpha_2 S^i_{(2)}
\right) 
\end{equation}
with $1/\alpha_k= 4 N_f^2 S_{(k)}(S_{(k)} + 1)(2 S_{(k)} + 1) /
3$ scales like $\sim e^{4N_f\alpha x}\sim D^{-4N_f\alpha}$ and
is relevant at small energy scales (low temperatures) $D\to 0$.
Therefore, any 'composite' fixed point is unstable.

Now we proceed by proving that the 'irreducible fixed point'
$\v^i \sim S^i$ is also unstable for $S\geq1$. To prove this it
is enough to find a relevant operator in the space of the
$\dv^i$'s. A detailed discussion of the construction of such an
operator is given in Ref.~\onlinecite{ZarPRL}, here we only give
its explicit form:
\begin{eqnarray}
{\cal O}_{\rm rel}& \sim & \dalpha (2N_f) \{
-{3\over 2} [ \tau^x (S^z S^x + S^x S^z) + \tau^y (S^z S^y +
S^y S^z)] \nonumber \\ 
& +&  \tau^z ({ S^x}^2 +  {S^y}^2 - 2 {S^z}^2 ) \} \; .
\label{eq:N_f:O_rel}
\end{eqnarray}
Substituting this expression into the scaling equations we find
that ${\cal O}_{\rm rel}$ scales like $\sim T^{- 8 N_f \alpha}$
and is relevant.

It is obvious from Eq.(\ref{eq:N_f:O_rel})
that the above considerations break down for an $S=1/2$ orbital
electron spin, since for
$S=1/2$ the operators appearing in (\ref{eq:N_f:O_rel}) vanish
identically as a 
consequence of the special properties of the $S=1/2$ spin
algebra. 
Thus, we can conclude that any low temperature fixed point of
Eq.(\ref{eq:N_f:nllscaling}) which is composed from several
spins or corresponds to 
an $S \geq 1 $ spin representation  is unstable.

\vskip0.4truecm
{ \noindent \it 2. Stability analysis of the $S_e=1/2$ fixed point}
\vskip0.4truecm

We now prove that this fixed point is really stable. 
To examine the stability we chose a basis in which the $\v^i$'s
at the fixed point can be written  in the blockmatrix form
\begin{equation}
(\v^i)_{\rm fp} ={1\over 2 N_f} 
\left(\matrix{{\tens \sigma}^i & \0 \cr
\0 & \0 \cr}\right)\; ,
\label{eq:N_f:v^i_0} 
\end{equation}
where ${\tens \sigma}^i$ ($i=x,y,z$) denote the Pauli matrices.  For the sake
of simplicity we assume that only a finite but arbitrarily large
number $N_o$ of orbital channels are considered, thus the
$\v^i$'s appearing in 
Eq.(\ref{eq:N_f:v^i_0}) are $N_o\times N_o$ matrices. Then the
deviation of the 
$\v^i$'s from their fixed point values can be written in the form
\begin{equation}
\delta\v^i = 
\left(\begin{array}{cc}
\etam^i & \t^i \\
{\t^i}^+ & {\tens M}^i 
\end{array}\right)
\; ,
\label{eq:fpdevi}
\end{equation}
where $\etam^i$, ${\tens M}^i$, $\t^i$ and ${\t^i}^+$ denote $2 \times 2$,
$(N_o-2)\times (N_o-2)$, $2 \times (N_o -2)$ and \break $(N_o-2) \times
2$ matrices, respectively. The matrices $\etam^i$ and ${\tens M}^i$ are Hermitian
while $\t^i$ and ${\t^i}^+$ are Hermitian conjugates
of each-other. The linearization of the scaling equations is lengthy but 
straightforward and one obtains the following set of equations:
\begin{eqnarray}
{d {\tens M}^i\over dx} &=& -{2\over N_f} {\tens M}^i\; ,  
\label{eq:N_f:mscaling} \\
{d\etam^i\over dx} &= & - {i\over N_f} \sum_{j,k}{ \epsilon^{ijk} }
({\tens \sigma}^j\etam^k + \etam^j{\tens \sigma}^k) - {2\over
N_f} \etam^i - {\tens \sigma}^i 
{1\over N_f} \sum_{j\not= i}{ Tr\{{\tens \sigma}^j\etam^j\}}
\nonumber \\
& + & \sum_{j\not= i}{{\tens \sigma}^j {1\over 2 N_f}Tr\{{\tens \sigma}^i\etam^j} +
{\tens \sigma}^j\etam^i\} \; ,
\label{eq:N_f:roscaling} \\
{d\t^i\over dx} & =& - {i\over N_f}
\sum_{j,k}\epsilon^{ijk}{\tens \sigma}^j\t^k - {2\over N_f} \t^i
 \; .\label{eq:N_f:tscaling}
\end{eqnarray}
The matrices ${\t^i}^+$ satisfy the Hermitian conjugate of
Eq.(\ref{eq:N_f:tscaling}).

The detailed stability analysis of the linearized scaling
equations is rather tedious.\cite{ZarPRL}
However, due to the complete decoupling of
Eqs.(\ref{eq:N_f:mscaling}), (\ref{eq:N_f:roscaling}), and 
(\ref{eq:N_f:tscaling}) all the scaling exponents can be
calculated exactly. Eqs.~(\ref{eq:N_f:mscaling} ---
\ref{eq:N_f:tscaling}) have an
infinite number of marginal solutions remaining unscaled under
the linear approximation. A thorough analysis reveals that these marginal
operators do not affect the $SU(2)$ spin algebra of the matrices
$(\v^i)_{\rm fp}$ and they correspond to the rotations of the pseudospin
of the TLS or the two-dimensional electronic orbital subspace,
where the $SU(2)$ algebra is realized.\cite{ZarPRL}
All the other operators scale like $\sim e^{-\exponent x} \sim
(D/ T_K)^\exponent$ to zero as $D\to0$ with an exponent $\exponent>0$ and are
irrelevant at low energies. 

\begin{figure}[htb]
\begin{center}
\epsfxsize=6.5cm
\hskip0.1pt\epsfbox{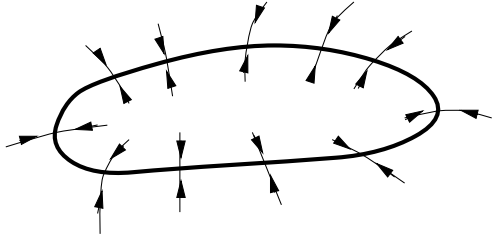} \\
\end{center}
\vskip0.3truecm
\caption{\label{fig:manifold} Sketch of the attractive $S_e=1/2$ manifold
embedded into the space of the couplings $\v^i$ ($i=1,2,3$). The
$S_e=1/2$ fixed manifold is represented by a heavy continuous
line while the scaling trajectories are indicated by arrows. A
marginal perturbation corresponds to moving along the heavy
continous line.}
\end{figure}

Accordingly, the $S_e=1/2$ 'fixed point' of the scaling
equation is rather an attractive 'fixed manifold' embedded in
the manifold of the general couplings $\v^i$. This situation is 
sketched in Fig.~\ref{fig:manifold}. The unitary transformations
of the $S_e=1/2$ 
spin algebra are associated to trajectories lying in this
attractive fixed manifold and correspond to zero exponents.
The number of leading irrelevant operators is
also infinite. These leading irrelevant operators associated
to the exponent $\exponent = 2/N_f$ are given by the following
expressions: 
\begin{eqnarray} 
{\cal O}_M & \sim & \sum_i \tau^i {\tens M}^i \; , 
\label{eq:N_f:O_M} \\
{\cal O}_J & \sim & \sum_i J^i \tau^i {\tens \sigma}^i \; , 
\label{eq:N_f:O_J} \\
{\cal O}_Q & \sim & \left ( \sum_i Q^i \tau^i\right) {\tens \sigma}^0 \; , 
\label{eq:N_f:O_Q} 
\end{eqnarray}
where the $Q^i$'s and $J^i$'s denote arbitrary small constants,
and ${\tens \sigma}^0$ is the unit matrix acting in the $2\times 2$ block
of the matrices in Eq.(\ref{eq:N_f:v^i_0}). 
The operator ${\cal O}_M$ appears due to the presence of the
other orbital channels of the conduction electrons. The second
operator, ${\cal O}_J$ corresponds to the leading irrelevant
operator of an anisotropic $N_f$ channel Kondo
problem.\cite{Gan,AL,Ye1} Finally, the third operator describes
the splitting of the TLS generated by the two dominant electron
channels coupled to the heavy particle.

\begin{figure}[htb]
\parbox{11cm}{
\hfill
\parbox{5cm}{
\epsfxsize=4.9cm
\hskip0.1pt \epsfbox{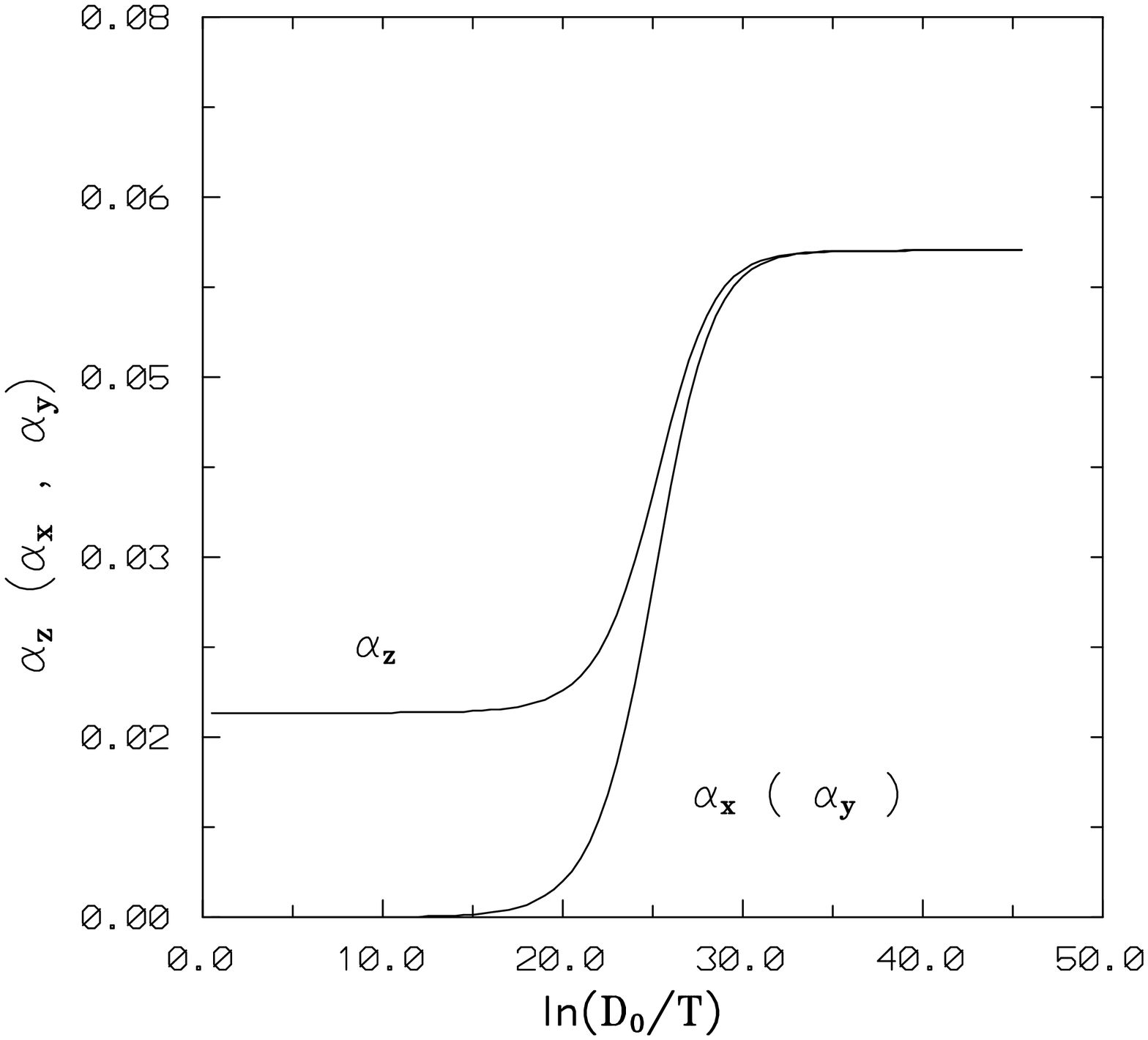} \hfill
\begin{center}
 Fig.~{\protect{\ref{fig:alpha}}}a
\end{center}}
\hfill
\parbox{5cm}{
\epsfxsize=4.9cm
\hfill
\hskip0.1pt\epsfbox{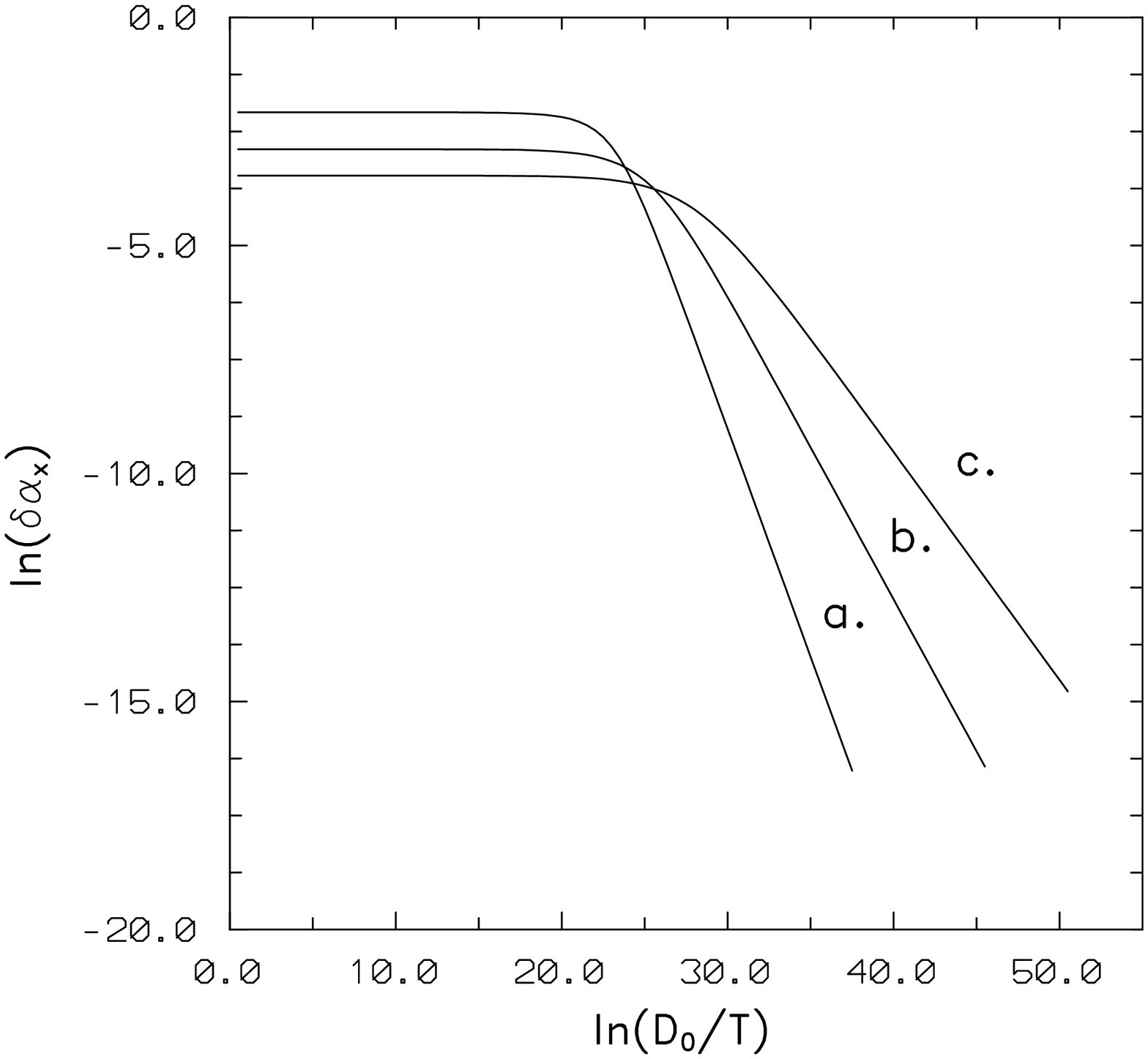} \hfill
\begin{center}
Fig.~{\protect{\ref{fig:alpha}}}b 
\end{center}
}} \hfill
\vskip0.3truecm
\caption{\label{fig:alpha} The scaling trajectory of the matrix norms
$\alpha_i= Tr\{{\v^i}^2 \}$ ($i=x,y,z$) for the $N_f=3$ case 
(Fig.~{\protect{\ref{fig:alpha}}}a) and their relaxation 
to their fixed point value (Fig.~{\protect{\ref{fig:alpha}}}b)
for $N_f=2$ (curve {\bf a.}), $N_f=3$ (curve {\bf b.}) and
$N_f=4$ (curve {\bf c.}) in a logarithmic scale.}
\end{figure}

The statements above have also been tested numerically.  In
Fig.~\ref{fig:alpha} we show the scaling of the norm
$\alpha_i = Tr\{(\v^i)^2\}$. The initial values of the couplings
have been estimated by assuming a screened Coulomb interaction
between the conduction electrons and the
TLS.\cite{VladZaw,ZarVlad} As one can see these scale to their fixed
point value $\alpha = {1\over 2 N_f^2}$ with the power law
dependence $\sim D^{2/N_f}$ in agreement with Eqs.~(\ref{eq:N_f:alpha}) 
and the scaling dimension of the leading irrelevant operators.
The Kondo effect can be identified as the breakdown of the curves.

\subsection{Scaling analysis up to the order $1/N_F^2$}
\label{ss:1/n_f^2}

\begin{figure}[t]
\epsfxsize=\hsize
\epsfbox{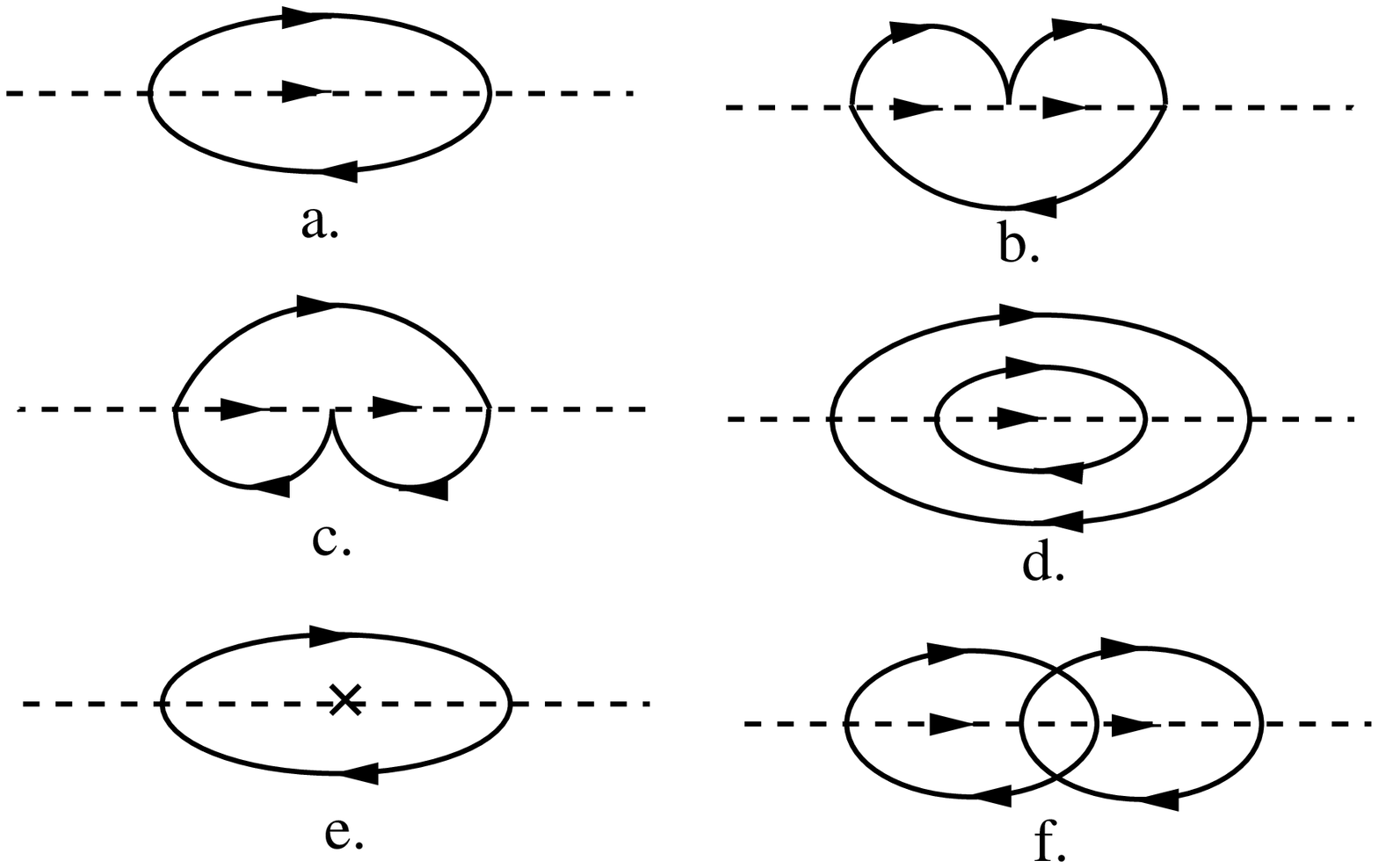}
\vskip0.7truecm
\caption{\label{fig:se_diagrams}
The pseudofermion self-energy corrections up to the order $\sim 1/N_f^2$.
The dashed and the continuous lines denote the
pseudofermion and the conduction electron propagators, respectively.
The cross indicates the contribution of the counterterm, which must be
calculated up to the order $\sim 1/N_f$.}
\end{figure}

\begin{figure}[t]
\epsfxsize=\hsize
\epsfbox{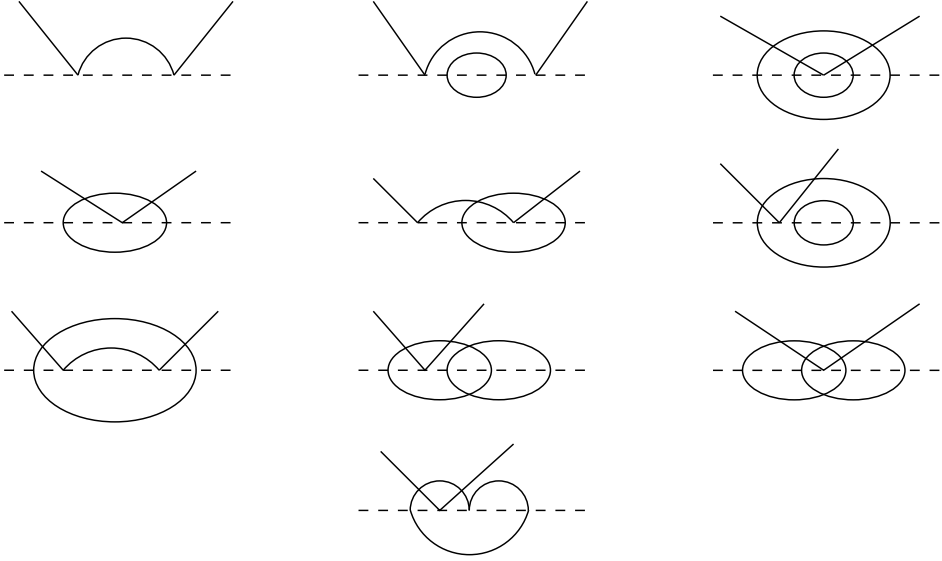}
\vskip0.7truecm
\caption{\label{fig:ve_diagrams}
The vertex corrections up to the order $\sim 1/N_f^3$.
For the sake of simplicity only diagrams without counterterm correction are 
shown. The missing diagrams can be generated by reversing the
pseudofermion and the electron lines.}
\end{figure}

To obtain the $1/N_f^2$ order scaling equations one has to
determine the vertex function and the pseudofermion self-energy
up to the orders $1/N_f^3$ and $1/N_f^2$, respectively.
The corresponding self-energy and vertex diagrams 
are shown in Figs.~\ref{fig:se_diagrams} and
\ref{fig:ve_diagrams}. They can, in principle, be calculated
easily using Abrikosov's pseudofermion technique,
however, there are some technicalities which are
crucial to obtain the correct results. 

The first poblem one has to face is the appearance of spurious
divergencies in 
the perturbation series.  The second order self-energy diagram,
Fig.~\ref{fig:se_diagrams}.a contains a logarithmic contribution but it also
gives a constant term $\delta\lambda= - 2 \; D\; \ln2 \; N_f{\rm Tr}\{
\v^i \v^i\}$ which renormalizes the chemical potential ('mass')
of the pseudofermions. One knows from quantum field theory that
such terms proportional to $D$ are artificial and can be scaled
out by introducing (up to second order) a counterterm in the Hamiltonian 
\begin{equation}
H_{\rm count}= -\delta\lambda \sum_\alpha b^+_\alpha b_\alpha\; .
\end{equation}
This counterterm is then canceling the spurious divergencies
in the different diagrams, and guarantees that the pole of the
pseudofermion Green's function remains unshifted. It has to be
determined order by order up to the desired order accuracy in
the perturbation theory. The inner loop in the self-energy
diagram of Fig.~\ref{fig:se_diagrams}.d, e.g., contains a 
constant part which results in an artificial contribution to
this diagram. This artificial contribution is then canceled by
the diagram \ref{fig:se_diagrams}.e.  If one uses this counterterm procedure
then the Abrikosov projection \cite{Abrikosov} of the $\sum b^+_\alpha
b_\alpha=1$ subspace should be carried out with the original 
$\lambda$ parameter in the Hamiltonian.

Another crucial problem is the separation of the different
logarithmic terms in the perturbation series. In the
perturbative expansion each term of the perturbation series can
be expanded in terms of the 
polinomials $\omega^n \ln^m \left( \omega / D \right)$ where
$\omega$ represents the relevant energy variable.
In a similar, way the total vertex and self-energy functions can
also be expanded in terms of such polinomials, and they can be
written schematically as
\begin{eqnarray}
\Sigma(\omega) &=& \sum_{n,m} \Sigma_{nm} \; \omega^n \ln^m \left( \omega / D
\right) \; , \\
\Gamma(\omega) &=& \sum_{n,m} \Gamma_{nm} \; \omega^n \ln^m \left(
\omega / D \right) 
\;,
\label{eq:expansion}
\end{eqnarray}
where the coefficients are some complicated matrix functions of
the couplings. Put in another way, the renormalization group
hypothesis Eqs.~(\ref{eq:RG_1}--\ref{eq:RG_3}) assumes that one
can change $D\to D^\prime$ and $\v^i 
\to {\v^\prime}^i$ in such a way that the expansion above
remains invariant up to a constant multiplicative factor. This
holds to {\it all the terms} appearing in the expansion, which
are all, we stress again, well-defined functions of the
couplings.  If one consideres the   $\Gamma_{02} \ln^2(\omega /D)$ vertex corrections,
e.g., rescaling $D$ one generates terms of the type $\sim
\ln(\omega/D) \ln(D/D^\prime)$, that renormalize only 
coefficients like $\Gamma_{01}$. 
The scaling equations can be,
of course, obtained equally well from the comparison of the
terms $\Gamma_{01}$ and $\Gamma_{02}$. These latters are,
however, much more difficult to calculate than the coefficients
$\Gamma_{00}$ and $\Gamma_{01}$. Therefore the easiest
way to generate the scaling equations is to collect the $\sim
\ln( D/ \omega)$ vertex corrections only. The difficulty
appearing now is to {\it separate} such contributions. The 
contribution of the  self energy diagram Fig.~\ref{fig:se_diagrams}.d
together with the regularizing counterterm contribution
 \ref{fig:se_diagrams}.e, e.g., can be calculated as
\begin{equation}
\Sigma_{\protect{\ref{fig:se_diagrams}}.d+e} = - \omega\;\left( -{1\over 2} \ln^2 {D\over
- \omega}+ (\ln2 -2) \ln {D\over -\omega} \right) N_f^2 \left( 
{\rm Tr} \{ \v^i \v^i\} \right)^2 + {\rm cnst.}\;,
\end{equation}
and it contains both $\omega \ln^2\omega$ and $\omega \ln\omega$
contributions from which only the second one contributes to the
wave function renormalization factor $Z_{p}$ in Eq.~(\ref{eq:RG_2})
and thus the scaling equations. This 
example shows, that in order to get the right scaling equations
one has to determine all the subleading contributions very
carefully. It also demonstrates, that Eqs.~(\ref{eq:RG_1}--\ref{eq:RG_2}) and
(\ref{eq:expansion}) must be taken {\it very seriously}.

Apart from these subtilities the calculation is straightforward
and after a rather tedious  calculation the inverse pseudo\-fermion Green's
function can be expressed as ($T=0$): 
\begin{eqnarray}
{\cal G}^{-1} &=& \omega \bigl\{ 1+ (1-\ln2)N_fO^{jj} +
\ln(D/\omega) \bigl[N_fO^{jj} + 12 N_f\beta \nonumber
\\&-& (5-3\ln2) N_f^2O^{jj}O^{kk}-(4\ln2-6)N_f^2O^{jk}O^{kj}\bigr]\bigr\}
\;, \label{eq:Gps}
\end{eqnarray}
where $O^{ij}={\rm Tr}(\v^i\v^j)$, $\beta=-i\,{\rm Tr}(\v^x\v^y\v^z-
\v^z\v^y\v^x)$ and
a summation must be carried out over the repeated indices.
Only the leading term proportional to $\ln(D/\omega)$ is given
explicitly in Eq.~(\ref{eq:Gps}) and we have taken
into account that the couplings are of the order of $\sim 1/N_f$.
The leading terms of the vertex function are given by
\begin{eqnarray}
\Gamma^i &= &\v^i-\ln2\,N_f\bigl(2O^{ij}\v^j-O^{jj}\v^i\bigr)
-\ln(D/\omega)\bigl[
2i\epsilon^{ijk}\v^j\v^k-N_f\bigl(2O^{ij}\v^j-O^{jj}\v^i\bigr)
\nonumber\\
&-& 4(2-\ln2)N_f i\epsilon^{jkl}O^{ij}\v^k\v^l 
+ 2\ln2\,N_f i\epsilon^{ijk}O^{ll}\v^j\v^k+
(2+5\ln2)N_f^2O^{jj}O^{kk}\v^i 
\nonumber\\
&-& (8+12\ln2)N_f^2O^{kk}O^{ij}\v^j + (8+12\ln2)N_f^2O^{ij}O^{jk}\v^k 
\nonumber \\ &-&
 (2+4\ln2)N_f^2O^{jk}O^{kj}\v^i \bigr] \;, \label{eq:gamma}
\end{eqnarray}
where $\epsilon^{ijk}$ denotes the Levi-Civita symbol.

Having determined the vertex function and the pseudofermion
Green's function we can generate the scaling equations by
plugging them into Eqs.~(\ref{eq:RG_1}--\ref{eq:RG_2}) in a selfconsistent way
and we finally obtain
\begin{eqnarray}
{d\v^i\over dx}&=&2i\epsilon^{ijk}\v^j\v^k+2N_f(O^{jj}\v^i-O^{ij}\v^j)-
8(1-\ln2)N_f i\epsilon^{jkl}O^{ij}\v^k\v^l
\nonumber \\ & +&16\ln2\,N_f\beta \v^i+
8N_f^2O^{ij}O^{jk}\v^k-8N_f^2O^{kk}O^{ij}\v^j
\;, \label{eq:scaling_N_f^2}
\end{eqnarray}
\noindent
where the scaling variable $x=\ln(D/D^\prime)$ has been used.
One can easily check
that with the substitution $\v^i\to j \sigma^i / 4$ these
equations simplify to the scaling equation derived
by Gan {\it et al.} for the $N_f$-channel Kondo model.\cite{Gan}

The stability analysis follows exactly the same lines as in the
previous Section. The only stable fixed
point of Eq.~(\ref{eq:scaling_N_f^2}) is identical with the fixed point
of the $N_f$-channel Kondo model:\cite{Gan}
\begin{equation}
v_{\rm fp}^i ={1\over 2 N_f}\left( 1 -{2\ln 2 \over N_f}\right)
\left(\matrix{{\tens\sigma}^i & 0 \cr 0 & 0 \cr}\right)\;,
\label{eq:fixpoint_N_f^2}
\end{equation}
where the Pauli matrices ${\tens\sigma}^i$ are acting in a two-dimensional
subspace in the electronic orbital space. In order to
determine the leading irrelevant operators one has to linearize 
the scaling equations around the fixed point (\ref{eq:fixpoint_N_f^2}).
For this purpose it is convenient to write the deviation
from the fixed point again in the form (\ref{eq:fpdevi}) and then the
linearized equations become: 
\begin{eqnarray}
&&{d \M^i\over dx} = - \exponent\, \M^i\;,
\label{eq:N_f^2:mscaling} \\&&
{d{\tens \varrho}^i\over dx} = - {i\over 2}\exponent \sum_{j,k}{ \epsilon^{ijk} }
({\tens\sigma}^j{\tens \varrho}^k + {\tens \varrho}^j{\tens\sigma}^k) - \exponent\,{\tens \varrho}^i
\label{eq:N_f^2:roscaling} \\&&\kern1.7pc
- {1\over 2}\exponent\, {\tens\sigma}^i \sum_{j\not= i}{ {\rm
Tr}\{{\tens\sigma}^j{\tens \varrho}^j\}}
+ {1\over 4}\exponent \sum_{j\not= i}{{\tens\sigma}^j \,{\rm
Tr}\{{\tens\sigma}^i{\tens \varrho}^j} +
{\tens\sigma}^j{\tens \varrho}^i\} \;,
\nonumber \\&&
{d\t^i\over dx} = -{i\over 2}\exponent
\sum_{j,k}\epsilon^{ijk}{\tens\sigma}^j\t^k - \exponent\, \t^i
 \;,\label{eq:N_f^2:tscaling}
\end{eqnarray}
where the critical exponent $\exponent = (2-4/N_f)/N_f$ is characteristic
to the stable fixed point. These equations have the same
structure as Eqs.~(\ref{eq:N_f:mscaling}--\ref{eq:N_f:tscaling}) the only difference is the appearance 
of the exponent $\exponent$ instead of the factor $2/N_f$.
Therefore all the conclusions concerning the operator content of
the fixed point remain in principle unchanged.
It is worth mentioning that, while the fixed
point couplings (\ref{eq:fixpoint_N_f^2}) contain a factor $\ln 2$
reminiscent of the special band structure used, there is no
similar factor in the expression of the critical exponent
$\exponent$, which is universal.

Similarly to the $1/N_f$-order case, there exist an infinite
number of leading irrelevant operators, in contrast to the 
multichannel Kondo model.
Theoretically, these new leading irrelevant operators could
change the low temperature properties of the model compared to
the Kondo model.  However, calculations of different measurable
quantities indicate that while they give a contribution to the
physical quantities, they do not change their critical exponent.
For the free energy, for example, we find a $\sim T^{2\exponent+1}$
behavior, while the scattering rate of the conduction electrons
scales as $\sim T^\exponent$.  With elementary considerations
one can also derive the scaling of the 'impurity magnetization',
$M_\Delta = \partial F_{\rm imp}/\partial \Delta$ and 
the 'impurity susceptibility' $\chi_\Delta = \partial^2 F_{\rm
imp}/\partial \Delta^2$ as a function of the splitting $\Delta$ 
and the temperature $T$. All these results are in agreement
with the exact ones obtained  for the $N_f$-channel Kondo model
discussed in the next Section.\cite{AL,Sacr_Schlott_rev} 
However, it would be interesting
to find some measurable quantity (like the Wilson
ratio,\cite{Wilson} for example) where the presence of these
operators is manifest. 

Naturally, the $1/N_f$ expansion breaks down in the physical
limit, $N_f\to 2$.  However, the structures of the leading
irrelevant operators being independent of the value of $N_f$ we
think that --- similarly to the $N_f$-channel Kondo model
\cite{Wilson} --- they are correctly given by
Eqs.~(\ref{eq:N_f:O_M}), (\ref{eq:N_f:O_J}), and (\ref{eq:N_f:O_Q}).
Therefore we think that the properties of a physical TLS with
$N_f=2$ could be determined by conformal field theory methods
\cite{AL} using a two-channel Kondo model, where these
leading operators are also included as perturbations around the
fixed point. We also expect that the scaling exponent of these
operators remains degenerate even for $N_f\to 2$ and that for
$N_f=2$ they coincide with the exact Kondo exponent,
$\exponent_{\rm exact} = 2 / (N_f+2)$.\cite{AL,Sacr_Schlott_rev}

\subsection{Analogy with the 2-channel Kondo model and scaling
behavior} 
\label{ss:nfl}

In the previous Sections we have basically established a mapping of the  
TLS model to the $N_f$-channel Kondo model. We have shown that at
the fixed point the couplings $\v^i$ take the particulary simple
form (\ref{eq:fixpoint_N_f^2}), $v^i_{nn^\prime} = v
\sigma^i_{nn^\prime}$, where the indices $n$ and $n^\prime$ 
label the two relevant orbital (angular momentum) channels.
Thus the TLS effective Hamiltonian can be written as
\begin{equation}
H_{\rm int}^{\rm eff} = v 
\sum_{\textstyle{\epsilon,n,\epsilon^\prime,n',s\atop
i,\alpha,\alpha'}} a^\dag_{\epsilon ns}b^\dag_\alpha
\sigma^i_{nn'}\tau^i_{\alpha\alpha'} b_{\alpha'}
a_{\epsilon^\prime n's} \;. \label{eq:H_eff}
\end{equation}
As the calculation in Sec.~\ref{sec:nll} demonstrates, the TLS
Hamiltonian becomes isotropical in the couplings $\v^x$, $\v^y$,
and $\v^z$ even above $T_K$. Moreover, we have shown 
in the previous Subsections that the
leading irrelevant operators of the TLS have the same scaling dimension
as the operator (\ref{eq:H_eff}). Therefore we conclude that the
effective Hamiltonian (\ref{eq:H_eff}) is adequate to describe
the TLS at temperatures around and below $T_K$.

Now the main observation is that the Hamiltonian
(\ref{eq:H_eff}) is formally equivalent to 
the one  of the $N_f$-channel Kondo model\cite{NozBland}
\begin{equation}
H_{\rm Kondo} = J \sum_{f=1}^{N_f}
\sum_{\epsilon,\epsilon^\prime,\sigma,\sigma^\prime,i} S^i
\sigma^i_{\sigma,\sigma^\prime} 
a^+_{\epsilon\sigma;f} a_{\epsilon^\prime \sigma^\prime;f}\; ,
\end{equation}
where now the $1/2$ inpurity spin is coupled to the real 
magnetic spins ($\sigma=\pm$) of $N_f$ independent electron
channels via an antiferromagnetic exchange interaction. The
correspondence between the two models is explained in
Table~I. 
\begin{figure}
\begin{center}
\parbox{11.5cm}{
\begin{tabular}{ccc}
\hline\hline
multichannel Kondo model  & \phantom{m} & TLS model \\
\hline
$H_{\rm i} =  J \sum S^i
\sigma^i_{\sigma,\sigma^\prime} a^+_{\epsilon\sigma;f} 
a_{\epsilon^\prime \sigma^\prime;f}$ && 
$H_{\rm i}^{\rm eff} = v 
\sum  a^\dag_{\epsilon ns}b^\dag_\alpha
\sigma^i_{nn'}\tau^i_{\alpha\alpha'} b_{\alpha'}
a_{\epsilon^\prime n's}$ \\ 
impurity spin $\;\; S^i$ && TLS pseudospin $\;\; \tau^i$ \\
electron spin  $\sigma$ && electrons' angular momentum $n=s,p,..$ \\
flavor $f=1,..,N_f$ && electrons' {\it real} spin $s=1,..,N_f$\\
local magnetic field $h$ && TLS splitting $\Delta$ \\
\hline\hline
\end{tabular}
\vskip0.5truecm
{\small TABLE I. Correspondence between the multichannel
Kondo and the TLS model.}}
\end{center}
\end{figure}
The multichannel Kondo model has already been studied by a
variety of methods and its properties are 
well-understood.\cite{Bickers,Andrei,Wiegman,Sacr_Schlott_rev,AL,2CK_NRG,VladZimZaw,FabrNozGog,EK}
Therefore our analogy between the two models is particularily
useful to understand the behavior of a TLS-model in details
and translate the exact results  obtained for the multichannel Kondo
model to the TLS case.

\begin{figure} 
\begin{center}
\epsfxsize=9cm 
\hskip0.1pt\epsfbox{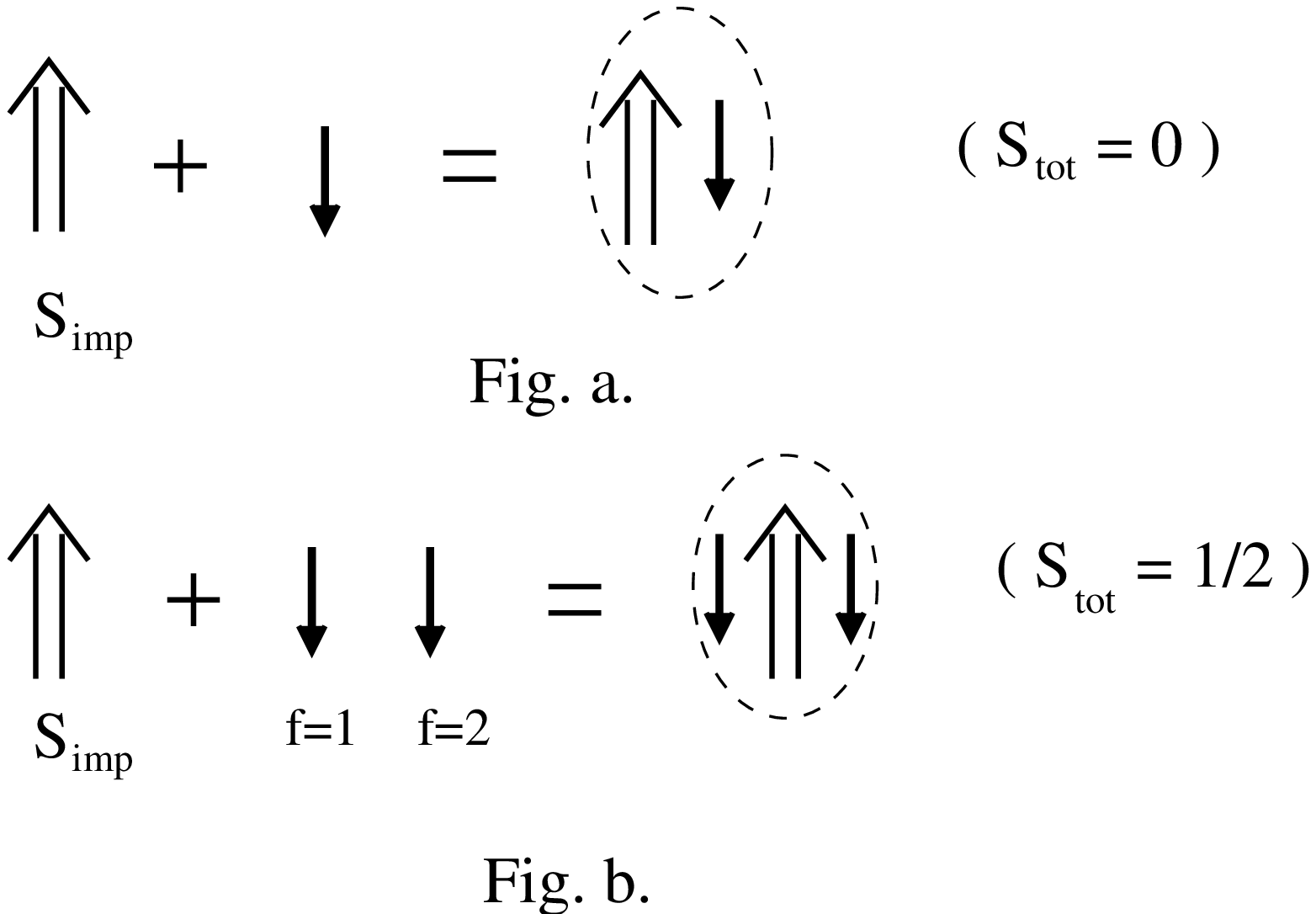}  \\
\end{center}
\vskip0.7truecm  
\caption{\label{fig:fusion} Comparison of the single channel Kondo
model and the two-channel Kondo model. In the two-channel Kondo model case the impurity spin is
overcompensated by the conduction electrons.} 
\end{figure}

For the sake of simplicity let us discuss first the case, $N_f=2$.
As we discussed above the physical TLS model is mapped to the
two-channel Kondo model, which, in contrast to the usual
single channel Fermi liquid Kondo model, belongs to the class of
{\it overcompensated} spin 
models and exhibits a non-Fermi liquid behavior. The origin of
the non-Fermi liquid behavior can easily be understood 
as follows. In the single channel case,
as the temperature is lowered below $T_K$, a conduction electron
is bound to the impurity spin by the antiferromagnetic
interaction, and they form together a spin singlet state (see
Fig.~\ref{fig:fusion}.a). This
composit singlet without inner degree of freedom 
is a 'dead body' for the rest of the electrons, which
are only slightly scattered by it, and therefore the ground
state of the system is properly described by a Fermi liquid
theory.  

On the other hand in the two-channel Kondo model around $T_K$ {\it both} channels
couple to the impurity spin antiferromagnetically and therefore,
as shown in Fig.~\ref{fig:fusion}.b, they form a bound state with an
effective spin $1/2$ pointing to the opposite direction as the 
impurity spin. A careful analysis shows that the resulting
bound object is coupled again antiferromagnetically to the rest
of the conduction electrons, and generates a new Kondo effect.\cite{NozBland}
Thus, in this picture the non-Fermi liquid behavior can be
viewed as a never-ending series of Kondo efects at different energy
scales. 
The fact that the conduction electrons
cannot screen the impurity spin in the multichannel Kondo
model is reflected in the 
appearance of a nonzero residual entropy due to the degeneracy
of the ground state:\cite{NozBland,Andrei,Wiegman,Sacr_Schlott_rev}
\begin{equation}
\Delta S_{\rm imp}^{\rm (2CKM)} = k_B {1\over 2}\ln 2 \;.
\label{eq:DS}
\end{equation}
Eq.~(\ref{eq:DS})  also gives the correct entropy for the TLS model
in the temperature range $T^*<T<T_K$, $T^*$ being the freezing
temperature.\cite{Andrei_crossover} 
However, for a real TLS the splitting acts like a local
magnetic field in the two-channel Kondo model, and splits up the degeneracy of the
ground state. therefore the splitting drives finally the TLS
towards a Fermi liquid state and  the residual entropy vanishes due to
the presence of the splitting as $T\to 0$:
\begin{equation}
\Delta S_{\rm imp}^{\rm (TLS)} = 0 \;.
\end{equation}

From the theory of the two-channel Kondo model we know that the specific heat and
the different susceptibilities show logarithmic
anomalies\cite{Andrei,Wiegman,Sacr_Schlott_rev} and 
that the resistivity scales like $T^{1/2}$ below $T_K$.\cite{AL}
These results can be translated for the TLS case without
difficulty keeping in mind the analogy (Table~I.) of the different
quantities.
The behavior of the corresponding physical quantities for the TLS
model is summerized in Table~II. both for the cases
$N_f=2$ and $N_f>2$, where the relevant energy variables,
$\Delta$ and $T$, are assumed to be larger than the freezing
temperature $T^*$ but smaller than $T_K$. 
\begin{figure}
\begin{center}
\begin{tabular}{c|cccc}
\hline\hline
  && $N_f>2$  && $N_f=2$ \\
\hline
$\chi_{\rm imp}^\Delta(\Delta,T=0)$ &\phantom{MM}& $\sim\left(\Delta \over
T_K \right)^{2/N_f\; -1}$ &\phantom{MMM} &  $\sim\ln \left(\Delta 
/ T_K \right)$
 \\
$\chi_{\rm imp}^\Delta(\Delta=0,T)$ && $\sim\left(T\over
T_K \right)^{(2-N_f)/(N_f+2)}$ &&  $\sim\ln \left(T 
/ T_K \right)$
 \\
$c_{\rm imp}/T$ && $\sim\left(T \over
T_K \right)^{(2-N_f)/(N_f+2)}$ &&  $\sim\ln \left(T 
/ T_K \right)$
 \\
$\Delta R_{\rm imp}$ && $\sim \left({T\over T_K}\right)^{2 / ( 2 +
N_f)}$ && $\sim \left({T\over T_K}\right)^{1/2}$ \\
\hline\hline
\end{tabular}
\end{center}
\vskip0.5truecm
{\small TABLE II. Low temperature behavior of the impurity
'orbital susceptibility' $\chi_{\rm imp}^\Delta = \partial^2
F_{\rm imp}/\partial\Delta^2$,  specific heat
$c_{\rm imp}$ and resistivity $R_{\rm imp}$ of a TLS. The energy
scales, $T^*$, $T$, $\Delta$, are assumed to be smaller than the
Kondo energy $T_K$.}
\end{figure}
\section{Generalization to the case of an $M$-state system}
\label{sec:Mstate}
In the previous Section we discussed the special case of
two-level systems. We have seen that at small energy scales
these can be discribed by the two-channel Kondo model.  The TLS
model is appropriate to describe tunneling centers in amorphous
metals,\cite{amorphous} where the probability of having a
three-state system is much smaller than that of having a TLS.
However, it breaks down in systems, where the tunneling centers
are formed by some substitutional impurities, and the heavy particle is
tunneling between 3, 6 or 8 equivalent positions. Tipical
examples of such alloys are the narrow gap semiconductor
${\rm Pb}_{1-x}{\rm Ge}_x{\rm Te}$ or insulating ${\rm
K}_{1-x}{\rm Li}_{x}{\rm Cl}$
alloys.\cite{Fukuyama,MLS} Three-state systems have also been 
observed in MOSFET devices where most of the tunneling defects
are TLS's due to some random structure.\cite{Cobden}  Therefore
it is natural to ask what is the low temperature behavior of an
M-level system (MLS) which strongly interacts with the
conduction electrons.

To answer this question we first have to generalize our TLS
model. Assuming that the 
temperature (or the relevant energy scale) is low enough 
and thus the motion of the heavy particle is now restricted  to the lowest 
lying $M$ states corresponding to the $M$ spatial positions of 
the  heavy particle the TLS Hamiltonian (\ref{11}) will be replaced by
\begin{equation} 
H_{hp} = \sum_{i,j=1}^M b^+_i \Delta^{ij} b_j \; , 
\end{equation}  
where $b^+_i$ creates a heavy particle at site $i$ and $\Delta^{ij}$ 
is the tunneling amplitude between positions $i$ and $j$.  
If we assume that no external stress is present and that the 
$M$ positions are completely equivalent thus $\Delta^{ii}=0$. 
This assumption is, however not necessary for the considerations
below.

Similarlly to the TLS case the we consider a general electron
spin $s$ taking the values $s=1\ldots N_f$ and we shall develop
a $1/N_f$ expansion around the fixed point. The kinetic
part of the Hamiltonian remains unchanged and
the most general two-particle interaction generated by some
effective interaction (screened Coulomb interaction or a pseudopotential) 
between the heavy particle and the conduction electrons now takes the form: 
\begin{equation} 
H_{\rm int}= \sum_{\scriptstyle i,j,n,m  
\atop\scriptstyle 
\epsilon, \epsilon^\prime} b^+_i V_{nm}^{ij} b_j 
a^+_{\epsilon   n   s} a_{\epsilon^\prime   m   s}  
\; .
\end{equation}  
For the sake of
simplicity we also assume a constant density of states per
flavor $\varrho_0$ between the high- and low-energy cutoffs $D$
and $-D$, for all flavor numbers.  Naturally, both the couplings
$V^{ij}_{nm}$ 
and the tunneling amplitudes $\Delta^{ij}$ are connected by the
symmetry properties of the MLS which will be exploited later on.

In the MLS case the pseudofermion propagator ${\cal
G}^{ij}(\omega)$ and the 
vertex function ${\tens\Gamma}^{ij}(\omega)$ are  $M\times M$ matrices
in the pseudofermion indices. Calculating these functions 
in a perturbative way it turns out that they do not satisfy 
the  simple multiplicative renormalization group (RG) 
equations (\ref{eq:RG_1}---\ref{eq:RG_3}) and the following generalized 
renormalization group transformation must be used ($T=0$):\cite{ZarPRL}
\begin{eqnarray} {\cal G}(\omega, \v^\prime,  
\Delta^\prime, 
D^\prime) & = & A \; {\cal 
G}(\omega, \v,\Delta, D)\; A^+ \; , \nonumber \\  
{\tens\Gamma}(\omega, \v^\prime, \Delta^\prime, D^\prime) & = &  
[ A^+]^{-1}  
{\tens\Gamma}(\omega, \v,\Delta,D) \; A^{-1} \; , 
\label{eq:mrg}  
\end{eqnarray}  
where the matrix notations $\varrho_0 
V^{ij}_{mn}\rightarrow 
\v^{ij}\rightarrow \v$, $ \Gamma^{ij}_{mn}\rightarrow
{\tens\Gamma}^{ij}\rightarrow {\tens\Gamma}$, and 
$\Delta^{ij} \rightarrow \Delta $ have been introduced,
$D^\prime$ stands for the scaled bandwidth and $A$ is an
$M\times M$ matrix acting in the heavy particle indices.  Note that
$A=A(\v^\prime,  \Delta^\prime,  
D^\prime/D)$ is independent of the dynamical variable $\omega$. 

This transformation has the following properties:
\begin{itemize}  
\item[(i)] For $A^{ij}= Z^{1/2} \delta^{ij}$ it is 
identical with the usual multiplicative renormalization group transformation. 
\item[(ii)] The symmetry Eq.~(\ref{eq:mrg}) is generated by the 
'microscopical' transformation $\v \to (A^+)^{-1} \v A^{-1}$, 
$\Delta \to (A^+)^{-1} \Delta A^{-1}$, ${\cal G}_0 \to A 
{\cal G}_0 A^+$, ${\cal G}_0$ being the bare heavy particle Green's 
function. 
\item[(iii)] The transformations (\ref{eq:mrg}) form a 
semi-group. 
\item[(iv)] Eq.~(\ref{eq:mrg}) leaves the Hamiltonian  
Hermitian. 
\end{itemize} 

While for finite $D/D^\prime$ the matrix $A$ has a rather 
complicated structure, still there exist such physical 
quantities like the electronic scattering rate or the 
impurity free energy, which are invariant under the 
transformation (\ref{eq:mrg}) and can be calculated 
easily.  It is important to note that for an infinitesimal
change of $D$ the matrix $A$ can be chosen to be Hermitian and
Eq.~(\ref{eq:mrg}) can 
be cast in the form of a scaling equation for the 
dimensionless couplings $\v^{ij}$. 
 
In the following we assume for the sake of simplicity that the
relevant energy variable is $\omega$, i.e., $\omega\gg
|\Delta^{ij}|, T$.  In this case the inverse heavy particle propagator and
the vertex functions can be expressed in the next to leading
logarithmic order \cite{VladZaw} as 
\begin{eqnarray} 
({\cal G}^{-1})^{ij} & = & \omega \; \delta^{ij} -  
\Delta^{ij} +  N_f \ln {D\over \omega} \;   
(\delta^{ij}\;\omega \;{\rm Tr}\{ \v^{kl} \v^{lk} \} - {\rm  
Tr}\{ \v^{ik} \Delta^{kl}  
\v^{lj} \}) \nonumber \\ 
\varrho_0 {\tens\Gamma}^{ij} & = & \v^{ij} - \ln{D \over \omega}  
\;  
\Big( [\v^{ik},\v^{kj}] - N_f {\rm Tr} \{ \v^{ik}  \v^{lj}  
\}\v^{kl}  
\Big) \; ,
\label{eq:verex} 
\end{eqnarray} 
where $[\phantom{m},\phantom{m}]$ denotes the commutator.
 Plugging (\ref{eq:verex}) into Eq.~(\ref{eq:mrg}) one can 
easily generate the scaling equations in a selfconsistent  
way and after some  algebra one finds: 
\begin{eqnarray} 
{d\Delta^{ij}\over dx} & = & -{1\over 2} N_f \Big[ {\rm  
Tr}\{  
\v^{ik} \v^{kl} \} \Delta^{lj} + \Delta^{ik} 
{\rm Tr}\{ \v^{kl} \v^{lj} \}  - 2 {\rm Tr}\{ \v^{ik}  
\Delta^{kl}  
\v^{lj} \}  
\Big] \; \nonumber \\ 
{d\v^{ij}\over dx} & = & - [\v^{ik},\v^{kj}] + {1 \over 2}  
N_f  
\Big(2 {\rm Tr} \{ \v^{ik}  \v^{lj} \}\v^{kl} - {\rm Tr}\{  
\v^{ik}  
\v^{kl} \}  
\v^{lj} - \v^{ik} {\rm Tr}\{ \v^{kl} \v^{lj} \} \Big), 
\label{eq:MLSscaling} 
\end{eqnarray}  
where $x=\ln{(D_0/D)}$ denotes the scaling variable.  
It is important to note that in the next to 
leading logarithmic level the scaling of the $\v^{ij}$'s  is 
completely independent of the splittings $\Delta^{ij}$, 
while the scaling of these latters is driven by the 
couplings $\v^{ij}$. 
 
These scaling equations are very complicated and, apart 
from some particular cases, one
can solve them only numerically. However, to exploit the  
symmetry properties of the MLS it is useful first to  
introduce a site representation in the orbital  
indices of the conduction electrons. We do this
by taking some linear combinations  
of the most strongly scattered angular momentum channels  
and hybridize them using group theoretical methods.  
For a regular 3-state system in the $xy$ plane and a free
electron band, e.g., one can use the three orthogonal electron 
states: 
\begin{eqnarray} 
|\epsilon\; 1 \rangle& = & {1 \over \sqrt{3}} | \epsilon\; s\rangle +  
\sqrt{2\over 3} | \epsilon\; p_x\rangle\; , \nonumber \\ 
|\epsilon\; 2\rangle & = & {1 \over \sqrt{3}} | \epsilon\; s\rangle -  
{1\over \sqrt{ 6 }} | \epsilon\; p_x\rangle + {1\over \sqrt{ 2  
}} | \epsilon\; p_y\rangle\; ,  \\ 
|\epsilon\; 3\rangle & = & {1 \over \sqrt{3}} | \epsilon\; s\rangle -  
{1\over \sqrt{ 6 }} | \epsilon\; p_x\rangle - {1\over \sqrt{ 2  
}} | \epsilon\; p_y\rangle\; , \nonumber  
\end{eqnarray} 
where the states $|p_x\rangle$ and $|p_y\rangle$ are defined in the  
usual way from the angular momentum states $\langle {\bf r}|\epsilon\;  
l\; m\rangle = \int (d^3{\bf k}/(2\pi)^3) Y_{lm}({\hat{\bf  
k}})e^{i{\bf k r}} \delta(\epsilon -\epsilon_{\bf k})$ with $l=1$.
Working with only those electron states which are 
directed to  the impurity positions the $\v$ becomes
an $M^4$-dimensional tensor. However, the number of  
independent couplings is largely reduced by  symmetry.  
For a heavy particle tunneling between the six corners of a regular  
octahedron, e.g., the $6^4=1296$ couplings may be replaced by 32  
independent couplings which makes a numerical solution  
reasonably fast.  

\begin{figure} 
\begin{center}
\epsfxsize=7cm  
\hskip0.5cm\epsfbox{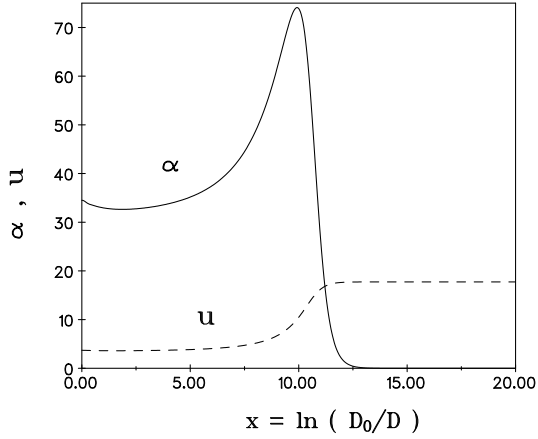}  \\
\end{center}
\vskip0.7truecm  
\caption{\label{fig:U's} Scaling of the norm of the  
dimensionless  
couplings, $u=\sum ||\v^{ij}||$ (dashed line), and of the  
algebra  
coefficient $\alpha$ (continuous line) for a 6-state 
system  
with $N_f=2$.  
The initial couplings have been chosen to be  
$v^{11}_{11}=0.8$, $v^{11}_{22} = 0.2$,  
$v^{11}_{66}=0.1$, $v^{11}_{12}=0.05$,  
$v^{11}_{16}=0.03$, $v^{12}_{21} =v^{12}_{12} = 0.0005$,  
$v^{12}_{11} = 0.005$, $v^{16}_{11} = 0.003$,  
$v^{16}_{61} =v^{16}_{16} = 0.0005$. The other nonzero  
couplings have been generated by symmetry  
transformations.} 
\end{figure} 

In Fig.~\ref{fig:U's} we show the typical scaling of the norm of
the dimensionless couplings, $u=\sum ||v^{ij}||$, 
(dashed line), where  the initial couplings have been estimated 
similarly to Refs.~\onlinecite{VladZaw,ZarZaw}. 
One can show similarly to the multichannel Kondo problem that
both the infinite and the weak coupling fixed points are unstable 
and the system scales to an intermediate strong coupling 
fixed point.\cite{NozBland,AL,2CK_NRG}  The Kondo energy 
can be defined as the crossover energy from the weak 
to strong coupling regimes:  $T_K = D_0 \; e^{-x_c}$, $x_c$ 
being the crossover value of the scaling parameter.  We foud
that for realistic initial parameters  this Kondo 
temperature can easily be found in the experimentally 
observable region, $T_K \sim \; 1- 10\;K$. 
 
Similarly to the TLS case we would like to determine the
properties of the MLS model in the regime $T^* < T,\omega \ll T_K$.
(Below the freezing temperature $T^*$ the dynamics of the MLS
becomes trivial in most cases.\cite{Moust}) To that purpose we have to determine the 
low-temperature effective model, i.e., the stable fixed points
of Eq.~(\ref{eq:MLSscaling}). Obviously, the scaling equation
has an infinite number of fixed points. Similarly to the TLS
case any representation of the $SU(M)$ Lie algebra will define a
possible fixed point. Trivially, the TLS fixed points, where 
two MLS states are completely decoupled from the other $M-2$
states, are fixed points of Eq.~(\ref{eq:MLSscaling}) as well.
Numerical investigations show that, apart from the fixed point
where the $\v^{ij}$'s realize the defining representation of the
$SU(M)$ Lie algebra, all these fixed points are unstable. 
The special fixed point corresponding to the defining representation
is equivalent to the the $SU(M)\times SU(N_f)$ Coqblin-Scrieffer
model.\cite{Coqblin}

In the following we shall show that the above-mentioned
Coqblin-Schrieffer model is really stable. To this
end we first remark that the operators 
$\delta^{ij}\sum_k \v^{kk}_{nm}$ are  
invariant under scaling. Therefore the $\v^{ij}$'s can be  
written as $\v^{ij}_{mn} = {\tilde v}^{ij}_{mn} + 
M^{ij}_{mn}$, where the matrix $M$ is a constant of motion
depending on the initial parameters and $\sum_i {\tilde
v}^{ii}_{nm} =0$. Then one can easily show  
that the right-hand side of Eq.~(\ref{eq:MLSscaling})  
disappears provided 
\begin{equation} 
{\tilde v}_0^{ij} = {1\over N_f} \left(\matrix{L^{ij} & 0 \cr 0 & 
0 
\cr}\right)\; 
, 
\label{eq:fp} 
\end{equation} 
where the $L^{ij}$'s are unitary equivalent to the  
generators of the $SU(M)$ Lie algebra, 
\begin{equation} 
[L^{ij},L^{kl}] = \delta^{il} L^{kj} - \delta^{kj}  
L^{il}\;. 
\label{eq:Lie}
\end{equation} 
with $L^{ij}_{nm} \sim  
\delta^i_m \delta^j_n - {1\over M} \delta^{ij}  
\delta_{nm}$.
To demonstrate that an MLS scales to this fixed point
in Fig.~\ref{fig:U's} we show the scaling of the 'algebra  
coefficient' 
$\alpha= \sum_{i,j,k,l} || N_f^2 [{\tilde v}^{ij},{\tilde 
v}^{kl}] - N_f  
\delta^{il} {\tilde v}^{kj} + N_f \delta^{kj} {\tilde  
v}^{il} || $, 
which measures in a natural way how well the fixed point  algebra
(\ref{eq:Lie}) is satisfied.
For $D <T_K$ (i.e. for  
$x>x_c$) the coefficient $\alpha$ scales to zero. Thus  
we conclude that {\it below the Kondo temperature an MLS scales
to the $SU(M)\times  SU(N_f)$ Coqblin-Schrieffer model, which is a  
non-Fermi-liquid  model} and has a different scaling behavior 
than the 2-channel Kondo model.\cite{AL,Cox_Ruckenstein}

To show that fixed point (\ref{eq:fp}) is stable and to analyze its 
operator content we follow the lines of the stability analysis
of the TLS fixed point and write the deviations of the couplings from
their fixed point value in form
\begin{equation}
\delta \v^{il} =
\left(\begin{array}{cc}
{\tens\varrho}^{ij} & \t^{ij} \\
(\t^{ji})^+ & \M^{ij}
\end{array}\right)
\;, \label{eq:deviation}
\end{equation}
where the couplings ${\tens \varrho}^{ij}$, $\t^{ij}$, and ${\tens\varrho}^{ij}$ are
$M\times M$, $M\times 
\infty$, and $\infty \times \infty$ matrices, respectively.
Like the TLS case the linearized equations for ${\tens \varrho}^{ij}$, 
$\t^{ij}$, and 
${\tens\varrho}^{ij}$ decouple completely,
\begin{eqnarray}
{d {\tens\varrho}^{il}\over dx} &=& {1\over N_f} \left( \delta^{ij}{\tens\varrho}^{kk} 
- 
M {\tens\varrho}^{il}\right)\;,
\label{eq:MLS:Mscaling} \\ 
 {d{\tens \varrho}^{il}\over dx} & =& - {1 \over N_f} 
\left([L^{ik},{\tens \varrho}^{kl}] + 
[{\tens \varrho}^{ik},L^{kl}] \right) 
+ {1\over 2N_f} \Bigl\{2\delta^{il}{\tens \varrho}^{kk} +  2 L^{jk}  {\rm 
Tr} \{{\tens \varrho}^{ij}  
L^{kl} + L^{ij}{\tens \varrho}^{kl}\} 
\nonumber \\ 
&-& 2 M {\tens \varrho}^{il}
-L^{ij}{\rm Tr}\{ {\tens \varrho}^{jk}L^{kl} + L^{jk} {\tens \varrho}^{kl}\} - 
{\rm Tr}\{{\tens \varrho}^{ij}L^{jk}
+L^{ij}{\tens \varrho}^{jk}\}L^{kl}\Bigr\}\label{eq:MLS:roscaling} \\
{d\t^{il}\over dx} &=& -{1 \over N_f} 
\left(L^{ik}\t^{kl} - L^{kl}\t^{ik}\right) + {1\over N_f}
\left(\delta^{il}\t^{kk} -  M \t^{il}\right)
 \;,\label{eq:MLS:tscaling}
\end{eqnarray}
and they can be solved exactly due to the simple structure of the 
$L^{ij}$'s. The linearized equations have an infinite number of
zero modes; a finite number  of them correspond to potential scattering while the others can 
be identified with the generators of the unitary transformations
connecting the different possible $M$-dimensional subspaces
where the $SU(M)$ Lie-algebra is realized. These 0-modes can be
shown, of course, to leave the Lie-algebra (\ref{eq:Lie}) 
invariant. All the other modes can be shown to be irrelevant, 
which proves the stability of the fixed point ({\ref{eq:fp}).

As we have pointed out in the previous Section, the universal
properties of a model are determined by the operator content of
its stable fixed points. Therefore, in order to determine the 
properties of the MLS in the region $T^*< T < T_K$, $T^*$ being
the freezing temperature, we have to determine the operator
content associated to the fixed point (\ref{eq:fp}). The analysis of
Eq.~(\ref{eq:MLS:tscaling}) shows that for $M\ge 3$ the leading
irrelevant operators can be written as 
\begin{equation}
{\cal O}_l \sim \left(\begin{array}{cc}
0 & C^{ij} \\
(C^{ji})^+ & 0
\end{array}\right)\;,
\label{eq:leading}
\end{equation}
where  the $C^{ij}$'s  satisfy 
$\sum_l(C^{kl}_{mn}-C^{ml}_{kn})=0$. These operators scale
like $\sim D^{\exponent_l}$ with $\exponent_l=(M-1)/N_f$, and they 
describe scattering between channels which are {\it not taken into
account} in the usual Coqblin-Schrieffer model. While they
dominate the thermodynamical quantities like the specific heat, 
e.g., which scales  as $c_{\rm imp}\sim T^{2\exponent_l}$ they do 
not contribute to the resistivity, which scales like $\sim T^{\exponent_{sl}}$
with $\exponent_{sl}=M/N_f$, and is determined by subleading 
operators  of the form
\begin{equation}{\cal O}_{sl} \sim \left(\begin{array}{cc}
Q^{ij}& 0 \\
0 & S^{ij}
\end{array}\right)\;,
\end{equation}
where the matrices $Q^{ij}$ and $S^{ij}$ satisfy $\sum 
Q^{ii}=\sum S^{ii}=0$ and $Q^{ij}_{mn} =Q^{ij}_{nm}$. It is
important to note that the operators (\ref{eq:leading}) do not 
exist in the TLS case ($M=2$), which explains why the low-energy 
properties of a TLS can be described by the two-channel Kondo
model.\cite{ZarVlad}

It is an open qestion whether the leading irrelevant operators found 
in the next to leading logarithmic approximation really exsist for 
the physical case $N_f=2$. Our results become, like the TLS
case, exact in the $N_f\to \infty $ limit. However, we expect
from conformal field theory\cite{AL_SU(M)} and noncrossing
approximation (NCA)  
results\cite{Cox_Ruckenstein} that the $1/N_f$ expansion breaks
down at $N_f=M$. Comparing the previous results with the
exponents of the  conformal field theory and recent Bethe Ansatz
calculations\cite{Andrei_M} $SU(M)\times SU(N_f)$ model we can
identify the exponents 
$M/N_f$ and $(M-1)/N_f$ with the exact ones $M/(M+N_f)$ and 
$(M-1)/(M+N_f)$. The expansion of the latters with respect to 
$1/N_f$ breaks down at $M=N_f$. Similarly, the NCA gives the
result that the $M>N_f$ and $M<N_f$ models behave in a different
way and that the $M=N_f$ case is marginal. We expect that,
similarly to the TLS case or the anisotropic Kondo model, the
{\it fixed point symmetry} remains the same even for 
$N_f<M$, however, the dimension of the different operators and
the operator  content might drastically change as we go  to
the region $M>N_f$. Therefore the role and the physical
interpretation of the new leading irrelevant operators in the
$N_f=2$ case is still rather obscur and needs some clarification.

Up to now we have ignored the effect of the MLS splitting.
These splittings, as mentioned before, can be taken into account
as infraread cutoffs and they finally stop the scaling towards
the $SU(M)\times SU(N_f)$ non-Fermi liquid fixed point at the freezing
temperature now defined by $T^* \sim \max_{\; i,j} |\Delta^{ij}(T^*)|$.
Much below this scale the MLS is frozen into its ground state. 
Usually this ground state is nondegenerate\cite{Moust} and
therefore, if $T^* > T_K$ then the non-Fermi liquid properties associated
to the fixed point (\ref{eq:fp}) can not be observed.
Therefore, the estimation of the splitting parameters and their
renormalization  under the scaling procedure is very important. 
\begin{figure} 
\begin{center}
\epsfxsize=7cm 
\hskip0.5cm\epsfbox{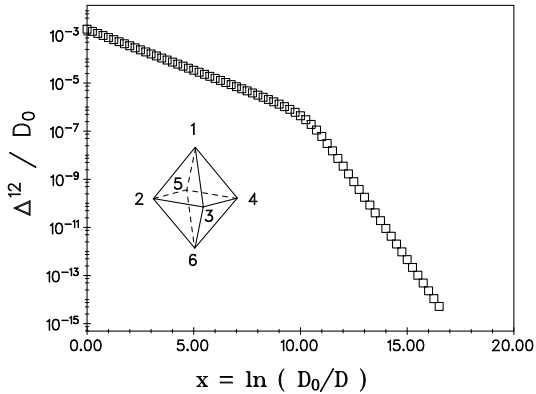}  \\
\end{center}
\vskip0.7truecm  
\caption{\label{fig:delta} Scaling of the  dimensionless  
hopping  
amplitude, $\Delta^{12}/D_0$ for the same 6-state system  
as in Fig.~1. Inset: Numbering of the  sites of the  
6-state system.} 
\end{figure} 
To estimate the renormalization of $\Delta^{ij}$ we solved
Eq.~(\ref{eq:MLSscaling}) numerically. As shown
in Fig.~\ref{fig:delta}, for realistic model parameters
$\Delta^{12}(T_K) / \Delta^{12}(D_0)$ can be  as small as $\sim
10^{-3}$, and therefore, even for very large splittings
$\Delta^{12}\sim 100 K$ the splitting is strongly reduced
and since $\Delta^{12}(T_K)\ll T_K$ the MLS can get easily
into the vicinity of the 2-channel Coqblin-Schrieffer fixed point.  
We stress at this point that, similarly to the TLS 
asymmetry energy $\Delta^z$, the renormalization of the diagonal
part of the pseudofermion (MLS) self-energy
$\sim(\Delta^{ii}-\Delta^{jj})$ is much smaller than that of the
tunneling amplitude $\Delta^{ij}$ ($i\not = j$).
However, for an MLS, (if there is no stress in the material, and
the concentration of MLS's is small), the previous ones  may be quite
small due to the symmetry of the MLS, and the formation of a
Kondo-like ground state with $T^* < T_K$, i.e., with
observable non-Fermi liquid behavior becomes quite reasonable. 
For a TLS in an amourphous material no such symmetry exsists and
the non-Fermi liquid properties are very probably smeard out in
the bulk system unless a very peculiar distribution of the TLS
splittings exists.

Unfortunately, metallic samples containing such highly
symmetrical MLS's due to some interstitials are not too
numerous. One example of them displaying a more or less unambigous Kondo
effect under preassure is the narrow gap semiconductor 
${\rm Pb}_{1-x}{\rm Ge}_x{\rm Te}$.\cite{Fukuyama} In this material the
relatively small $Ge^{2+}$ ions form 8-level systems and
interact with the conduction electrons.  Alas, this
material is a very complicated one and its preparation was, at
the time of the measurements, not well controlled neither. In these
samples the concentration of the MLS's is rather high and
their interaction probably results in the nonvanishing of the
difference of the MLS site energies, $\Delta^{ii}-\Delta^{jj}$,
which hinders the observation of the non-Fermi liquid properties.
 Moreover in the
${\rm Pb}_{1-x}{\rm Ge}_x{\rm Te}$ alloy the spin-orbit interaction is also
large\cite{} and may result in a cross-scattering between the
two spin-channels. This process also spoils the non-Fermi liquid
behavior. Thus, in order to observe the predicted non-Fermi
liquid property of an MLS one should find some better candidate.

One of the other possibilities is to use nanotechnology to
construct devices to realize the $SU(M)\times SU(2)$ models. 
Such a realization of the $SU(3)\times SU(2)$ model has been
suggested in Ref.~\onlinecite{Andrei_M} by means of a double dot
structure.

\section{Short review of new theoretical developments}
\label{sec:new}

In the previous Sections we mainly concentrated on the 
mapping of an MLS model to the $SU(M)\times SU(N_f)$
Coqblin-Schrieffer model by using the multiplicative renormalization group technique.  
These issues, however, constitute only a fraction of 
the presently available results for tunneling Kondo
impurities. The aim of the present Section is to review
in a concise way such techniques and results that lie out of the
scope of the present work. To get a deeper understanding of them
the interested  reader is referred to the original works.

\subsection{Path integral treatment}

One of the drowbacks of the multiplicative renormalization group treatment of the TLS (MLS)
model is that it is only reliable in the small coupling region
where the perturbative expansion is meaningful. On the other
hand, simple estimations show\cite{VladZaw} that for reasonable
TLS parameters the screening interaction $v^z$ might be quite
strong, $v^z\sim 0.4$, where the multiplicative renormalization group is already not applicable.
This discrepancy of the multiplicative renormalization group has partially been cured in
Ref.~\onlinecite{VladZimZaw} where a Yuval-Anderson type scaling
analysis\cite{YuvalAnd} has been applied to the TLS model.

In this approach, as a first step, one devides the Hamiltonian
into two parts, $H_{\rm noflip}$ and $H_{\rm flip}$, where all the
terms associated to the flip of the TLS ($v^x$, $v^y$,
$\Delta^x$, and $\Delta^y$) are grouped in the part $H_{\rm
flip}$. The main observation is that $H_{\rm noflip}$ contains
all the large parameters of the model and can be solved exactly. Therefore,
calculating the partition function $Z$ of the model, one can develop
a perturbation theory in the small TLS-flipping interactions.
Due to the relatively simple structure of $H_{\rm noflip}$ one
can integrate out the electronic degrees of freedom in each term
of the perturbation series. As a result of this procedure one is
left with an effective action for the TLS pseudospin only.

As usually, this effective action is fairly complicated and
containes logarithmic retarded interactions between the
different TLS flips. However, one can use it to generate renormalization group
equations even if $v^z$ is large, by requiring the invariance 
of the partition function under the renormalization group transformation. 
The prcedure is, however, very sensitive to the exact way of
elimination of the high energy degrees of freedom and needs
lots of care.\cite{e-hole} Several different elimination shemes exist
and some of them  leave the partition function invariant but not the 
different dynamical Green's functions. Another disadvantage is
that the expansion of $Z$ is {\it not systematic} and by construction
does not treat the different couplings on equal footing. Up to
now no-one succeded to derive the next to leading logarithmic
scaling equations with this method.

The TLS model has also been studied recently by a bosonization
technique combined with scaling analysis with somewhat different
results from the previous ones.\cite{MF}
In this method the electronic degrees of freedom are
represented by their bosonic charge (spin, flavor, and
spin-flavor) density fluctuations which makes the problem easier
to handle. Very probably most of the aformentioned differences arize from the
fact that in Refs.~\onlinecite{MF} such an elimination scheme has
been used that mixes together the TLS and conduction
electron degrees of freedom in an artificial way.\cite{e-hole} 
On the other hand, the bosonization procedure uses a very
special cutoff scheme and it is presently not clear for us
if the anomalous dimension of the different couplings is
correctly given by it at a finite phase shift for an arbitrary
process such as multi-electron scattering. 

The Anderson-Yuval technique has also been succesfully 
applied to the problem of the tunneling of a heavy particle on a
lattice.\cite{Vladar_latt} For the noncommutative
case, however, the scaling equations become very involved and little
physical  information can be obtained from them. This problem has also 
been studied using the multiplicative renormalization group
method in Ref.~\onlinecite{Zawa_latt}. 

\subsection{Role of the excited states}
\label{ss:excited}

In the previous Sections we simply ignored the excited states of
the TLS by the handwaving argument that they are completely
frozen out at the low temperatures we are interested in  and only the
lowest-lying nearly degenerate states must be kept. These
excited states, however, influence the low-energy behavior of
the TLS. Even if the temperature is low virtual processes
consisting of several 'assisted hoppings' to the excited states
like the one shown in Fig.~\ref{fig:excited} are allowed,
generating an effective assisted tunneling process from one well
to the other. Since the overlap of the states $|l \rangle $
($|r\rangle $) and $|ex\rangle$ is large compared to that of
$|l\rangle$ and $|r\rangle$, these processes cannot be neglected.
To take them into account one can easily generalize
the Hamiltonians (\ref{11}) and (\ref{eq:13}) to
\begin{eqnarray}
H_{\rm TLS} &=& \sum_i E_i b^+_i b_i \;, \nonumber \\
H_{\rm int} &=& \sum_{\textstyle{\epsilon,n,\epsilon^\prime,n' \atop
i,j, s}} a^\dag_{\epsilon ns}b^\dag_\alpha
V^{ij}_{nn'} b^+_i b_j a_{\epsilon^\prime n's}\;,
\end{eqnarray}
where now the operators $b^+_i$ create a state with the heavy
particle in the {\it exact} eigenstate of the double well
potential $|i\rangle$ with energy $E_i$, where $E_1=0\approx E_2 \ll E_3<E_4 <..$.

\begin{figure}[htb]
\parbox{12cm}{
\hfill
\parbox{5.5cm}{
\epsfxsize=5.5cm
\hfill\epsfbox{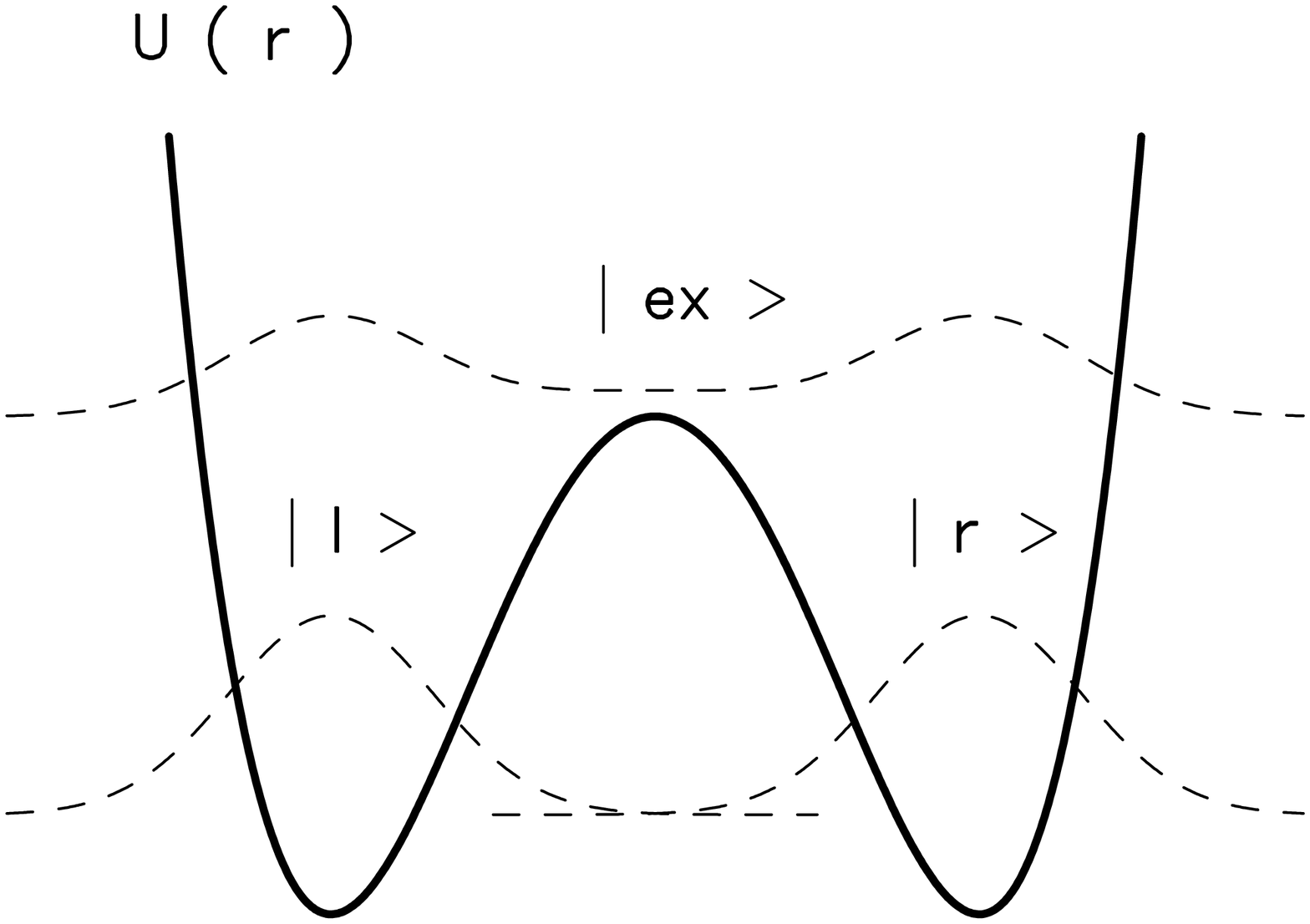}\hfill
\begin{center}
 Fig.~{\protect{\ref{fig:excited}}}a
\end{center}}
\hfill
\parbox{5.5cm}{
\epsfxsize=5.5cm
\hfill\epsfbox{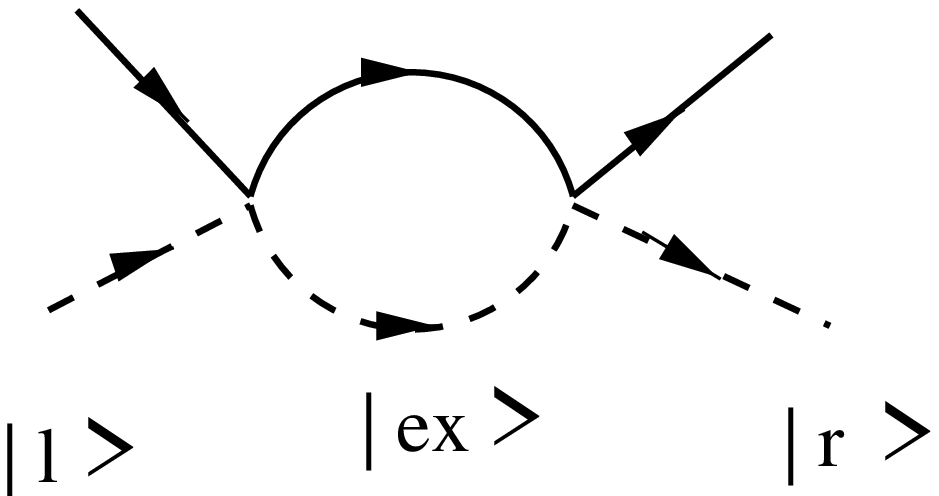}\hfill
\begin{center}
Fig.~{\protect{\ref{fig:excited}}}b 
\end{center}}\hfill}\hfill

\vskip0.3truecm
\begin{center}
\parbox{12.5cm}{
\caption{\label{fig:excited} 
Fig. a: Sketch of a double well potential and
the states $|r>$, $|l>$, and the first excited state $|ex>$.
Fig. b: A diagram generating the assisted tunneling via virtual
transitions to the excited states.}} 
\end{center}
\end{figure}
A careful estimation of the assisted hopping matrix elements and
a multistep scaling analysis\cite{ZarZaw} reveal that for a
typical TLS the presence of the excited states may increase the
Kondo temperature by a factor of $10-100$ thereby giving
$T_K\sim 1-10K$ which is in much better agreement with the
experimental results as our previous estimation (\ref{5.16}). 
It has also been shown in Ref.~\onlinecite{ZarZaw} that at low
temperatures the assisted hopping to the excited states can be
taken into account by the redefinition of the couplings 
$V^\mu_{nn^\prime}$ in Eq.~(\ref{eq:13}). Thus one can ignore
formally the excited states but then the TLS-electron couplings
$\v^\mu$ should be viewed as some effective 
couplings incorporating the influance of them as well.

\subsection{A TLS with spin}

To obtain a non-Fermi-liquid scaling of the TLS it has been
crucial that the scattering of the conduction electrons be
diagonal in ther {\it real } spin ($\sim$ flavor), $s=1,..,N_f$.
In this case the Hamiltonian possessed an additional
$SU(N_f)$ symmetry associated to the flavor degrees of freedom.
From the group theoretical point of view\cite{AL} it is this 
additional $SU(N_f)$ symmetry that is responsible for the 
non-Fermi liquid behavior of the model.
Therefore it is natural to ask what happens if the tunneling
particle has also a spin which is coupled to the conduction
electrons by a local exchange interaction $\sim J S^i
\sigma^i({\bf R})$, $\sigma^i({\bf R})$ being the local spin
density of the conduction electrons at the heavy particle's
location ${\bf R}$. 
In this case, in addition to the usual interaction
(\ref{eq:13}) an exchange interaction term appears of the form
\begin{equation}
\sum_{\scriptstyle i,\mu,s,s^\prime,n,n^\prime \atop
\scriptstyle \epsilon,\epsilon^\prime,\sigma,\sigma^\prime,\alpha,\alpha^\prime}
 J^\mu_{n n^\prime}
b^+_{\alpha s} \tau^\mu_{\alpha \alpha^\prime}S^i_{s s^\prime}
 b_{\alpha^\prime s^\prime}
(a^+_{\epsilon n \sigma} \sigma_{\sigma \sigma^\prime}^i
a_{\epsilon^\prime n^\prime \sigma^\prime})\;.
\label{eq:exch}
\end{equation}
This interaction obviously breaks the $SU(2)$ symmetry in the
electronic spin, and it mixes together the two electronic spin
channels. A simpler version of Eq.~(\ref{eq:exch}) has already
been studid in Ref.~\onlinecite{2CK_NRG} in the context of magnetic
impurities in crystal field.

A straightforward renormalization group study of the model Eq.~(\ref{eq:exch})
shows\cite{tspin} that, as one could guess it from the outset, the coupling
to the impurity spin always drives the system to a Fermi liquid
ground state even in the absence of the TLS splitting $\Delta$.
The nature of this ground state may, however, depend on the
initial parameters of the model. 

\begin{figure}[htb]
\parbox{12cm}{
\hfill
\parbox{5.5cm}{
\epsfxsize=5.5cm
\hfill\epsfbox{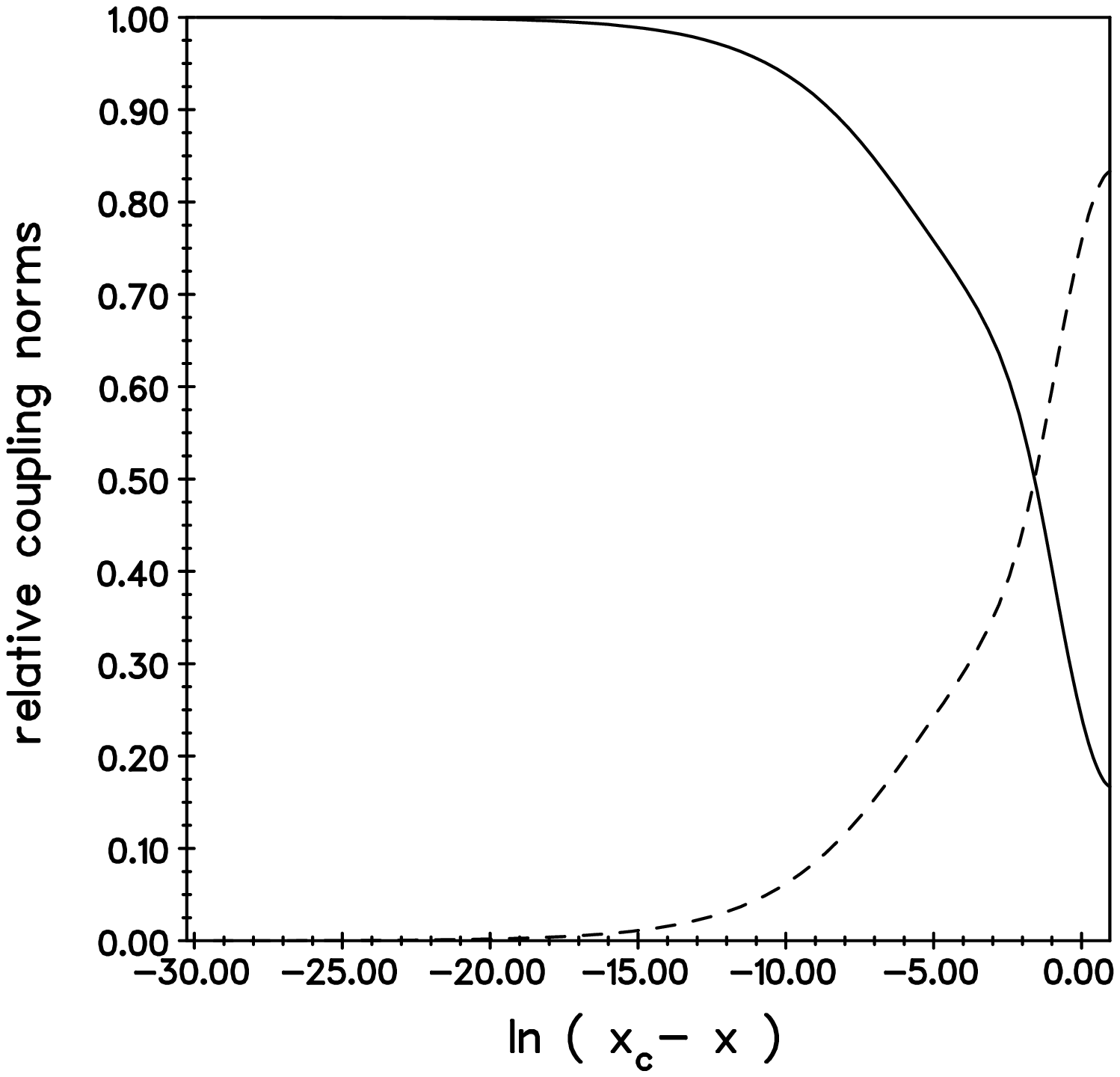}\hfill
\begin{center}
 Fig.~{\protect{\ref{fig:transition}}}a
\end{center}}
\hfill
\parbox{5.5cm}{
\epsfxsize=5.5cm
\hfill\epsfbox{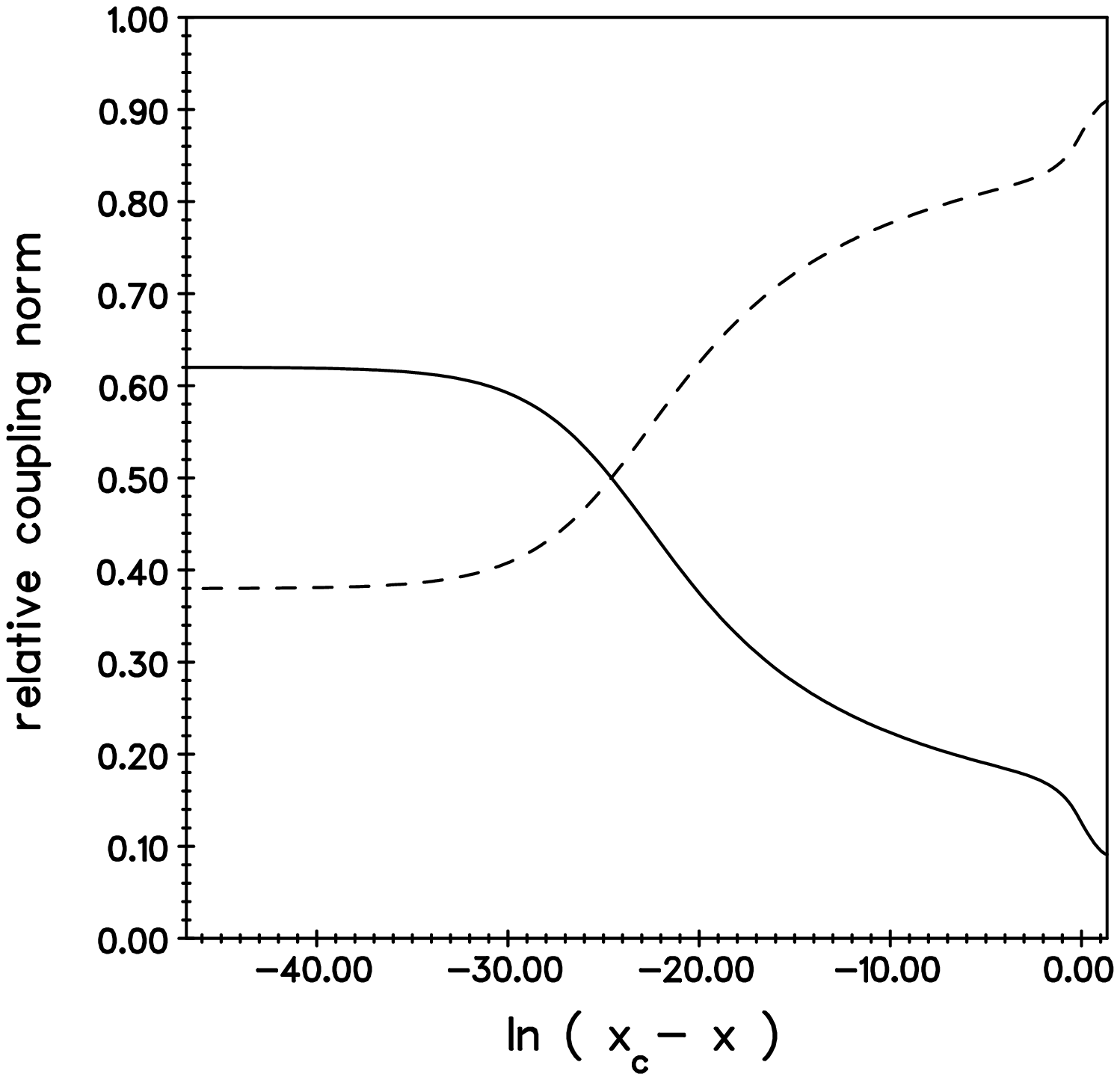}\hfil
\begin{center}
Fig.~{\protect{\ref{fig:transition}}}b 
\end{center}}\hfill}\hfill

\vskip0.3truecm
\begin{center}
\parbox{12.5cm}{
\caption{\label{fig:transition} 
The scaling trajectory of the relative norms $||v||/(||j|| +
||v||)$ (dashed line) and $||j||/(||j||+||v||)$ (solid line)
in a logarithmic scale for two different set of initial
couplings. The scaling trajectories go from right to left, and
$x_c=\ln(D_0/T_K)$.  In Fig.a the exchange interaction dominates the
fixed point behavior of the model. For the small exchange coupling case
in Fig.b. a different fixed point appears, characterized by a finite
universal ratio $||j||/||v||$. }} 
\end{center}
\end{figure}
In Fig.~\ref{fig:transition} the scaling of the relative norms of the
dimensionless couplings ${\tens j}^\mu$ and $\v^\mu$ is shown for two
sets of initial couplings in the leading logarithmic
approximation. In Fig.~\ref{fig:transition} the model scales to a
fixed point where the couplings $v^\mu$ can be neglected
compared to the exchange couplings. In this case the generalized
exchange coupling (\ref{eq:exch}) generates a dynamical
splitting for the TLS and therefore the orbital motion of the
TLS is frozen out at low temperatures, and the low-energy
behavior of the model can be appropriately described by a
spin $1/2$ single channel magnetic Kondo model.\cite{tspin}

On the other hand in Fig.~\ref{fig:transition}b we show another
situation where the ratio $||j||/||v||$ scales to a finite
value. In this case at the fixed point both the orbital and
magnetic degrees of freedom are coupled to the TLS in a
symmetrical way and the low-energy effective model can be shown
to be equivalent to the single channel $SU(4)$
Coqblin-Schrieffer model.

The tunneling spin problem has also been treated very recently 
using conformal field theory methods which resulted 
similar results to the present ones.\cite{Ye2}

\section{Review of the experimental situation}
\label{sec:exp}

Presently there are only a few systems where Kondo-like
anomalies attributed to TLS's or other tunneling centers have
been observed.

Logarithmic anomalies due to TLS's formed probably by dislocations
have been found in some amorphous systems like Al$_{1-x}$Mg$_x$
or Al$_{1-x}$Ge$_x$ \cite{dislocations} In these materials the typical 
splitting of the TLS's is large ($\sim 1-10K$) and a wide
distribution of the TLS splitting is present, therefore the
non-Fermi-liquid regime can never be observed. These anomalies
are not sensitive to magnetic field and they disappear upon
annealing. 

Another extensively studied system containing Kondo-like
tunneling centers is the alloy ${\rm Pb}_{1-x}{\rm Ge}_x {\rm Te}$ that we already
discussed in detail at the end of Section~\ref{sec:Mstate}. 
While in some circumstances an unambigous logarithmic
anomaly due to TLS's can be observed in this
material,\cite{Fukuyama}  it is 
too complex from the theoretical point of view and the
interplay of different fenomena (phonon softening, resonant
scattering from phonons, ferroelectric phase transitions) make
it very difficult to analyse. 

\begin{figure}
\vskip6cm
\caption{\label{fig:Ralph} Universal scaling behavior of 
the unannealed $Cu$ point contact spectrum from Ref.~
\protect{\onlinecite{Ralph_Ludwig}}}
\end{figure}
The most promising experiments concerning TLS Kondo impurities
have been performed on metallic point
contacts.\cite{Ralph_Buhr,Ralph_Ludwig} In these 
experiments only scattering on a few TLS's in the contact region
influences the signal of the device, and the differential
conductance of the point contact is directly proportional to the
temperature and energy dependent scattering rate of the
electrons on these TLS's,\cite{Yanson} $G(V,T)={dI\over dV}\sim
{1\over\tau(T,V)}$.  In these
measurements not only a logarithmic temperature and voltage
dependence at high energies has been observed  but  also 
a {\it universal scaling} $\sim [\max(T,V)]^{1/2}$ of the
conductance curves shown in Fig.~\ref{fig:Ralph} has been found.
The two-channel Kondo scaling curves calculated by noncrossing
approximation\cite{Kroha} and conformal field
theory\cite{JanLudwig} fit very well the observed scaling curves. 

Altshuler et al. suggested that a similar scaling behavior could
result from electron-electron interaction in localization
theory.\cite{Altshuler} However this explanation has several constituents
which are clearly in contradiction with the experimental
facts (an anomalously  short estimated mean free path being in contradiction
with the good resolution of the phonon peak, disappearance of
the signal with increased static disorder, magnetic field
independence of the signal in some
experiments\cite{Shashi}).\cite{Ralph_answer} We should mention,
however, that 
while one could see in some experiments the cutoff of the scaling by
some intrinsic splitting at low energies in complete agreement
with the TLS theory,\cite{Shashi} in others no violation of 
the universal scaling has been observed at $T\sim 1K$. This
should imply that TLS's with small renormalized splittings
$\Delta^*<1K$ are very likely to occur in the latter systems. 
This might be due to some special origin of the TLS's but it is
presently not well-understood.
\begin{figure} 
\begin{center}
\epsfxsize=7cm 
\hskip0.5cm\epsfbox{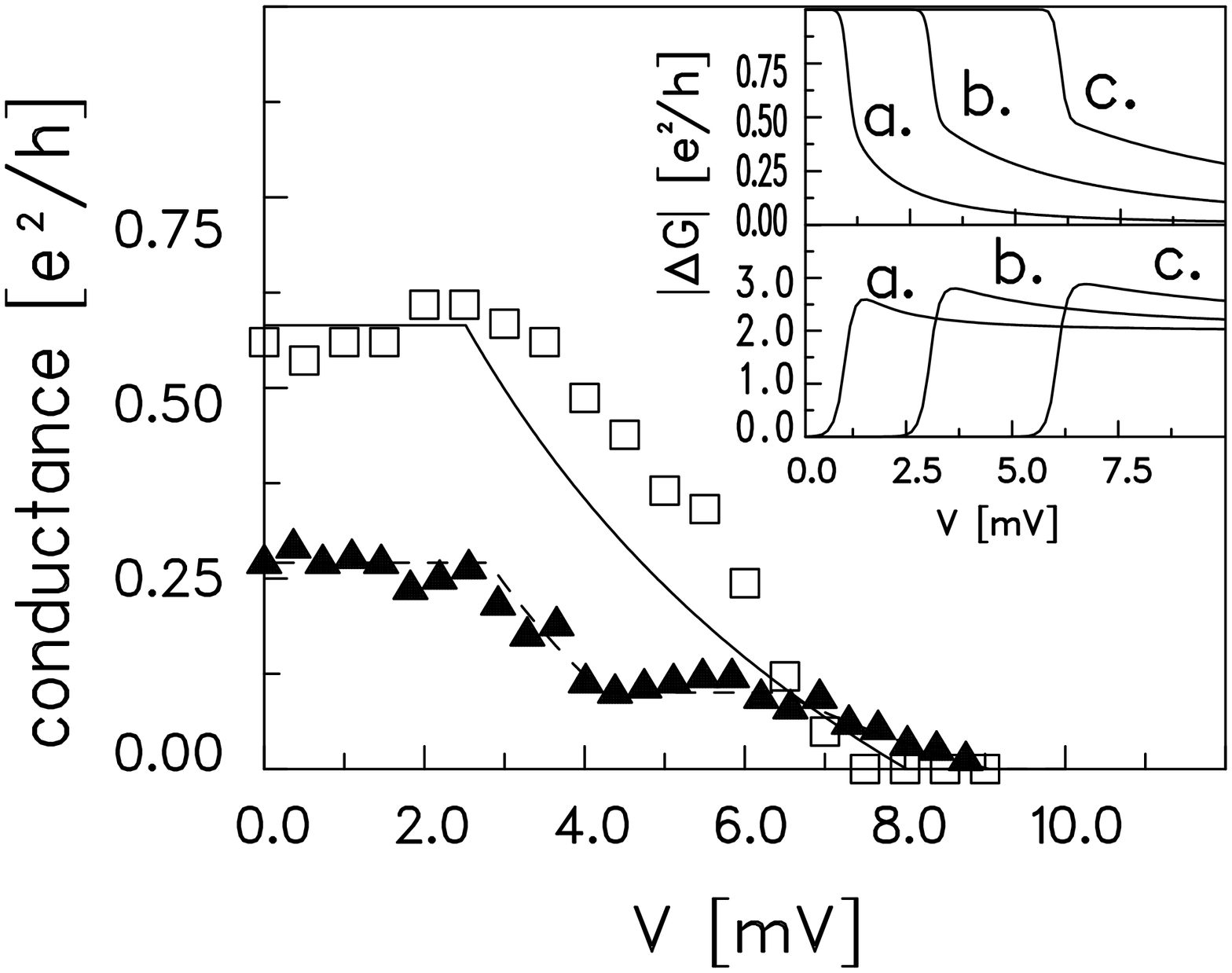}  \\
\end{center}
\vskip0.7truecm  
\caption{\label{fig:Keijsers} The squares and triangles 
correspond to $\Delta G (V)$ calculated from curve 3 in Fig.~2
and the curve of Fig.~4 of Ref.~\protect{\onlinecite{Keijsers}}.
The continuous curves are fits with the Kondo theory (the dashed
line has been fitted by scattering from two TLS's). The inset
shows curves calculated with the Kozub-Kulik theory\protect{\cite{Kozub}}
incompatible with the experimental results.}
\end{figure} 
Finally, we discuss the results of a very recent measurement on
amorphous break junctions by Keijsers et al.\cite{Keijsers}
giving a very nice 
possibility to observe directly a single noncommutative TLS. 
In these experiments Keijsers et al. found that the zero-bias
anomaly $G(V)$ ($R(V)$) of the point contact fluctuated slowly 
between two values, $G_1(V)$ and $G_2(V)$. 
Since the zero-bias anomaly is thought to be due to the fast
two-level systems 
present in the contact area these fluctuations can be explained 
in terms of a slow fluctuator that influances the parameters of
the fast TLS's.\cite{Keijsers} One can also argue that the
difference $\Delta G=G_1(V)-G_2(V)$ should be due to only a few
possibly only one fast TLS  with modified parameters.\cite{JanZarZaw}
Since it is the asymmetry parameter $\Delta^z$ of a TLS that is
most sensitive to external stress, it should be possible to fit 
the difference $\Delta G(V)$ by changing the splitting
parameters $\Delta_i \to \Delta_i^\prime$ for just a few TLS
Kondo curves.\cite{JanZarZaw}  Our fit to $\Delta G(V)$ is shown in
Fig.~\ref{fig:Keijsers}, and is in very good agreement with the 
experimental curves, thus supporting the existence of fast
Kondo-like two-level systems in amorphous metals.

\section{Conclusions}
\label{sec:conclusions}

In the present paper we tried to give a comprehensive and
somewhat detailed review of the present understanding of TLS
Kondo impurities. While we mainly concentrated on the
multiplicative renormalization group approach and to the mapping
of M-level systems to the much simpler $SU(M)\times SU(N_f)$
model we also attempted to review shortly other theoretical and
experimental results in the field. To understand these more in
detail the interested reader is adviced to consult the given
References directly.   Of course, this review is
far from being complete since the TLS literature is growing
very fast. The role of excited states is still not well
understood, very little is known about the influence of Kondo
TLS's on the other properties of the material and the properties
of M-level systems.

We hope that  we managed to convince the reader that tunneling 
Kondo impurities deserve further studies and provide very
promising realizations for the two-channel Kondo model, even
though the discussions are not settled  about them yet.

\acknowledgements
The authors would like to thank N. Andrei, D.L. Cox, Jan von
Delft, R. Keijsers, A. Moustakas, P. Nozi\`eres, A. Zawadowski, and G.T.
Zim\'anyi for helpful discussions.  G. Z. has been supported by 
the Magyary Zolt\'an Fellowship of the Hungarian Ministry of Education.
This research has been supported by the Hungarian Grants OTKA~T021228, 
OTKA~F016604, and OTKA~T017128.


\begin{references}
\bibitem{Moust} It has been suggested in Ref.~\protect{\onlinecite{MF}}
that the ground state of an MLS could be degenerate due to some
dynamical Jahn-Teller effect, this is, however not very likely 
to occur in reality.
\end{references}
\end{document}